\documentclass[preprint,showpacs,showkeys]{revtex4}  
\usepackage{epsf, epsfig, latexsym}
\usepackage{graphicx}
\usepackage{amssymb}
\usepackage{amsmath}
\usepackage{bm}
\usepackage{bbm}
\usepackage{dcolumn}
\usepackage{multirow}
\usepackage{pifont}
\usepackage{calc}
\begin{document}
\bibliographystyle{apsrev}
\setcounter{page}{0}
\title[]
{Gluon Propagation in Curved Space  }
\author{Su Kyeong \surname{Lee}, Eun-Joo \surname{Kim}}
\author{Jong Bum \surname{Choi}}
\email[]{jbchoi@jbnu.ac.kr}
\affiliation{Division of Science Education and Institute of Science Education,
Chonbuk National University, Jeonju 561-756, Korea}
%
\date[]{Received {\today}}
%
%
%
\begin{abstract}
%
%
In quantum chromodynamics (QCD), the vacuum structures are taken to be formed
by nonperturbative interactions between gluons.
The contributions of gluonic interactions can be parametrized by gluon condensates
and especially the dimension-2 condensate $\langle A_{\mu}^{2} \rangle$
can be used to generalize the $\theta$ vacuum.
In the generalized picture, each gauge fixed slice forms a curved surface around a hadron
and other strongly interacting objects,
and the propagation of a gluon has to be considered on this curved surface.
The gluon propagator turns out to be massive due to the curvature of the gauge fixed surface
implementing the possibility that the massless gauge bosons might provide
the majority of the hadronic mass and even behave as dark matter in cosmological scales.
\end{abstract}
%
%
%
%
\maketitle
%
%
%
%
\section{INTRODUCTION}
\label{sec1}
%

The strong interactions are taken to be mediated by gluons and the most peculiar
aspect of gluonic interactions is the nonperturbative self-interactions.
These nonperturbative self-interactions generate the problem of definition of vacuum state
which is the starting point of any calculation by using the properties of quantized fields such as propagators.
The vacuum states for strong interactions are known to be classified according to
their topological quantum numbers~\cite{R1}
and those classified states turn out to be mixed through instantons~\cite{R2}.
The condition that the vacuum state has to be stationary with respect to gauge transformations
leads to the definition of $\theta$ vacuum~\cite{R3} resulting from the undetermined angle $\theta$
appearing in the eigenvalue of gauge transformation. Because the number of possible values of
$\theta$ is infinite, the gluonic vacuum states can exist in infinitely different states physically
independent from each other. However, in ordinary picture, infinitely many vacuum states may generate
infinitely different vacuum expectation values which lead immediately to contradictory results
with experimental observations.

The solution to the problem of infinitely many vacuum states appeared as a by-product of the
consideration of in-hadron condensates~\cite{R4}.
The vacuum condensates~\cite{R5} are defined as the nonzero vacuum
expectation values of normal ordered products of field operators and are usually taken to be constant
with respect to space-time variables.
A well-known effect of vacuum condensates is the modification of quark and gluon propagators resulting
in the damping in propagation~\cite{R6}. The damping arises due to background fields residing within hadronic
scales and therefore the vacuum condensates have to be considered only within the scale of hadronic size.
In this way the idea of in-hadron condensates has been suggested and various examples are discussed
with the introduction of maximum wavelength of quarks and gluons confined in hadrons.
Although the in-hadron condensate idea resolved many problems, there existed intrinsic restrictions
concerning the boundary values of condensates if the condensate values are taken to be constant inside hadrons.
The only breakthrough to this restriction is to assume the variation of condensate values
resulting in the new picture of QCD vacuum~\cite{R7} composed by union of different gauge fixed slices.

The new picture of QCD vacuum generalizes the picture of $\theta$ vacuum with coordinate dependent
$\theta$ values~\cite{R8}. These coordinate dependences can be parametrized by the values of dimension-2
condensates~\cite{R9} and for a given value of dimension-2 condensate the set of points constitute
a curved surface around a hadron. Each surface represents one gauge fixed slice corresponding to
one $\theta$ vacuum and the infinitely many $\theta$ vacua can be drawn as
different surfaces around a hadron. Of course this picture can be applied to the more complicated
situations such as multi-quark states or nuclei and even to the extreme case of quark-gluon plasma.
In fact, the generalization of $\theta$ vacuum with coordinate dependences has been suggested as
one possible scenario~\cite{R10} to explain the strong CP problem in hot quark-gluon matter.
Recent experiments at RHIC and LHC confirmed the P-odd charge separation effects~\cite{R11} and
the generalized $\theta$ vacuum can be taken as the state
where the gluonic behaviors have to be analyzed.

Calculational method for the structures of gauge fixed slices has been developed by considering
topological spaces of gluonic domains~\cite{R12} which can be classified according to the numbers of quarks
and antiquarks inside each gluonic domain. For a given topological space, the changes of gluonic
domains are induced by quark pair creation or annihilation and the probability amplitude to have
a quark pair is taken to be proportional to the value of dimension-2 condensate at that point~\cite{R13}.
When these considerations are combined with the amplitude defined by nonlocal condensate between
two quarks, we can deduce functional form of the variation of dimension-2 condensate.
The surface formed by the points with the same value of dimension-2 condensate constitutes
one gauge fixed slice and we can draw infinitely many slices by changing the value of dimension-2
condensate. Thus the QCD vacuum around a hadron or other structures such as multi-quark states and nuclei
can be pictured by the union of each gauge slice formed by given value of dimension-2 condensate.

For a given QCD vacuum, we can try to deduce the Feynman rules and the most important quantity
is the gluon propagator. In order to deduce the form of gluon propagator we have to fix the
gauge and the quantization procedure has to be carried out on this gauge fixed slice.
But now the gauge fixed slices turn out to be curved surfaces and the quantization of gluons
has to be considered on the curved space.
The first effects of curved space are the changes of derivatives from the ordinary ones into
the covariant ones~\cite{R14}. These changes give rise to additional terms related to the curvature
of the space. In the simplest case the additional term can be interpreted as carrying out
the role of mass term in gluon propagator, and thus there appears the possibility
that the masslessness of gluons as gauge bosons can be changed naturally into
the massiveness of hadrons or other strongly interacting objects.
For cosmological scales these considerations may lead to
one possible candidate for dark matter~\cite{R15}.

In Sec. II, we review the procedure of generalization of $\theta$ vacuum,
and in Sec. III, we show the structures of gauge fixed slices.
In Sec. IV, we deduce the form of gluon propagator in curved space,
and in Sec. V, we suggest gluons as dark matter candidate.
The final section is devoted to summary and discussions.
%
%
%
\section{GENERALIZED $\theta$ VACUUM}
\label{sec2}
%
%
For atomic structure, the composing elements are nucleus and electrons and these elements
have definite masses with nonrelativistic movements. The vacuum state for an atom is taken
to be a state in which nothing exists. From this vacuum state we can count the number of
electrons and an atom is formed with these definite number of electrons and nucleus.
However, as we approach the nucleus it becomes impossible to count the number of electrons
because electron-positron pairs are continuously created and annihilated as the energy of
the probing particle goes beyond 1~MeV. Then the vacuum state for a nucleus cannot be a state
of nothing. It is a state of somethings that are created and annihilated continuously.
In order to describe physical quantities systematically we need to arrange the creation and
annihilation of particles in a definite manner and the resulting method is known as normal ordering.
In quantum field theories, the method of normal ordering is fully exploited to define physical quantities.

The situation changes once again when we consider nonperturbative self-interactions of gluons.
The normal ordering method is useful only when we can count the number of particles in a given state.
But due to the nonperturbative self-interactions the number of gluons in a state cannot be counted
appropriately. Therefore vacuum expectation values of normal ordered field operators
do not vanish for gluonic states and these are parametrized as vacuum condensates.
The original idea of condensate was applied only to gauge invariant forms of field operators
and the field operators were classified according to their dimensions.
However, after three loop calculations had been carried out~\cite{R16},
there appeared the necessity of introducing dimension-2 condensate
 which depended on the choice of gauge.
If something is dependent on the choice of gauge, it can be used to fix the gauge.
Since the quantization procedure of gluons has to be carried out with gauge fixing,
the slice with given value of dimension-2 condensate can be the appropriate place
where the quantization of gluons could be performed. The collections of these gauge
slices are intimately related to the generalization of $\theta$ vacuum.

The definition of $\theta$ vacuum starts from the instanton solution to minimize the
Yang-Mills action satisfying
\begin{equation}
  F_{\mu\nu} \equiv \partial_{\mu}A_{\nu} - \partial_{\nu}A_{\mu} + g[A_{\mu}, A_{\nu}] = 0.
\label{eq1}
\end{equation}
To induce the form of $A_{\mu}$ satisfying these equations, we consider the transformation properties
of $A_{\mu}$ at long distances and get
\begin{equation}
   A_{\mu}(x) = g^{-1}(x) \frac{\partial g(x)}{\partial x^{\mu}},
\label{eq2}
\end{equation}
where $g(x)$ represents the matrix element of the group of gauge transformations.
In Euclidean space with imaginary time, the set of points at infinity in four dimensional space forms
a large sphere $S^3$ and the above relation generates mappings between the points on $S^3$ and
the group elements $g(x)$. These mappings classify the gauge fields $A_{\mu}(x)$ into infinitely many
independent components. For a group element $g_{1}(\mathbf{x})$,
the gauge field $\mathbf{A}_{1}(\mathbf{x})$ for a vacuum is given by
\begin{equation}
   \mathbf{A}_{1}(\mathbf{x})= g^{-1}_{1}(\mathbf{x})\nabla g_{1}(\mathbf{x}).
\label{eq3}
\end{equation}
Representing the mapping corresponding to $n$ times repetition of $g_{1}(\mathbf{x})$ as
\begin{equation}
  g_{n}(\mathbf{x})= [g_{1}(\mathbf{x})]^{n},
\label{eq4}
\end{equation}
the gauge field given by
\begin{equation}
   \mathbf{A}_{n}(\mathbf{x})= g^{-1}_{n}(\mathbf{x})\nabla g_{n}(\mathbf{x})
\label{eq5}
\end{equation}
constitutes another vacuum state locally.
When we write the wave functional for this vacuum state as $\psi_{n}[\mathbf{A}]$,
the other vacuum state $\psi_{n+1}[\mathbf{A}]$ at $\mathbf{A}_{n+1}(\mathbf{x})$
can exist and these states can be mixed through tunnelling.
Then the true vacuum state can be written as
\begin{equation}
  \Psi[\mathbf{A}] = \sum_{n} C_{n}\psi_{n}[\mathbf{A}],
\label{eq6}
\end{equation}
and the coefficients $C_{n}$ can be determined from the condition that
$\Psi[\mathbf{A}]$ has to be stationary with respect to gauge transformations.
The final result is
\begin{equation}
  \Psi_{\theta}[\mathbf{A}] = \sum_{n} e^{in\theta}\psi_{n}[\mathbf{A}],
\label{eq7}
\end{equation}
and this state is called the $\theta$ vacuum.

Now the parameter $\theta$ remains as a free parameter.
For two different $\theta$ and $\theta^{\prime}$
the transition amplitude becomes
\begin{equation}
 \langle \theta^{\prime} |~e^{-Ht}~ |\theta  \rangle \xrightarrow{t \to \infty}\delta (\theta - \theta^{\prime} )
    \sum_{\nu} e^{-i\nu\theta} \int [DA_{\mu} \cdots]_{\nu}
    \exp \{-\int d^{4}x \mathcal{L}(A_{\mu} \cdots)\}.
\label{eq8}
\end{equation}
The exponential factor in front of the functional integral represents the change of vacuum winding number
and this factor can be inserted into the functional integral by changing the Lagrangian density as
\begin{equation}
 \mathcal{L} = \mbox{tr}\left[F_{\mu\nu}F^{\mu\nu}+\frac{\theta g^{2}}{8\pi^{2}}F_{\mu\nu}\tilde{F}^{\mu\nu}\right]
\label{eq9}
\end{equation}
in Minkowski space. The second term breaks the $P$ and $CP$ invariances and these effects
have been checked by the measurement of neutron electric dipole moment resulting in the limit
$\theta < 3 \times 10^{-10}$.
However, in recent experiments at RHIC and LHC, local P-odd effects are observed
implementing the possibility that topological fluctuations exist in hot quark-gluon matter.
Then there exist definite differences between the gluonic states of a neutron and
hot quark-gluon matter.
Because the differences are generated by instantons, we need to analyze the relationships
between the instanton contributions and the structures of QCD vacuum.
These relationships have been checked by considering the correlation function on a lattice
in one approach, and by using the instanton shape recognition procedure on the lattice in
another approach. From the comparison of two-point correlation function with the lattice
gluon propagator, $O(\frac{1}{k^2})$ correction to gluon propagator has been confirmed
indicating the existence of dimension-2 condensate $\langle A_{\mu}^{2} \rangle$~\cite{R17}.
By using instanton shape recognition procedure~\cite{R18} we can also confirm the presence of
dimension-2 condensate and the expected values are well explained in an instanton-gas
or an instanton-liquid picture.

\begin{figure}
\includegraphics[width=0.76\linewidth]{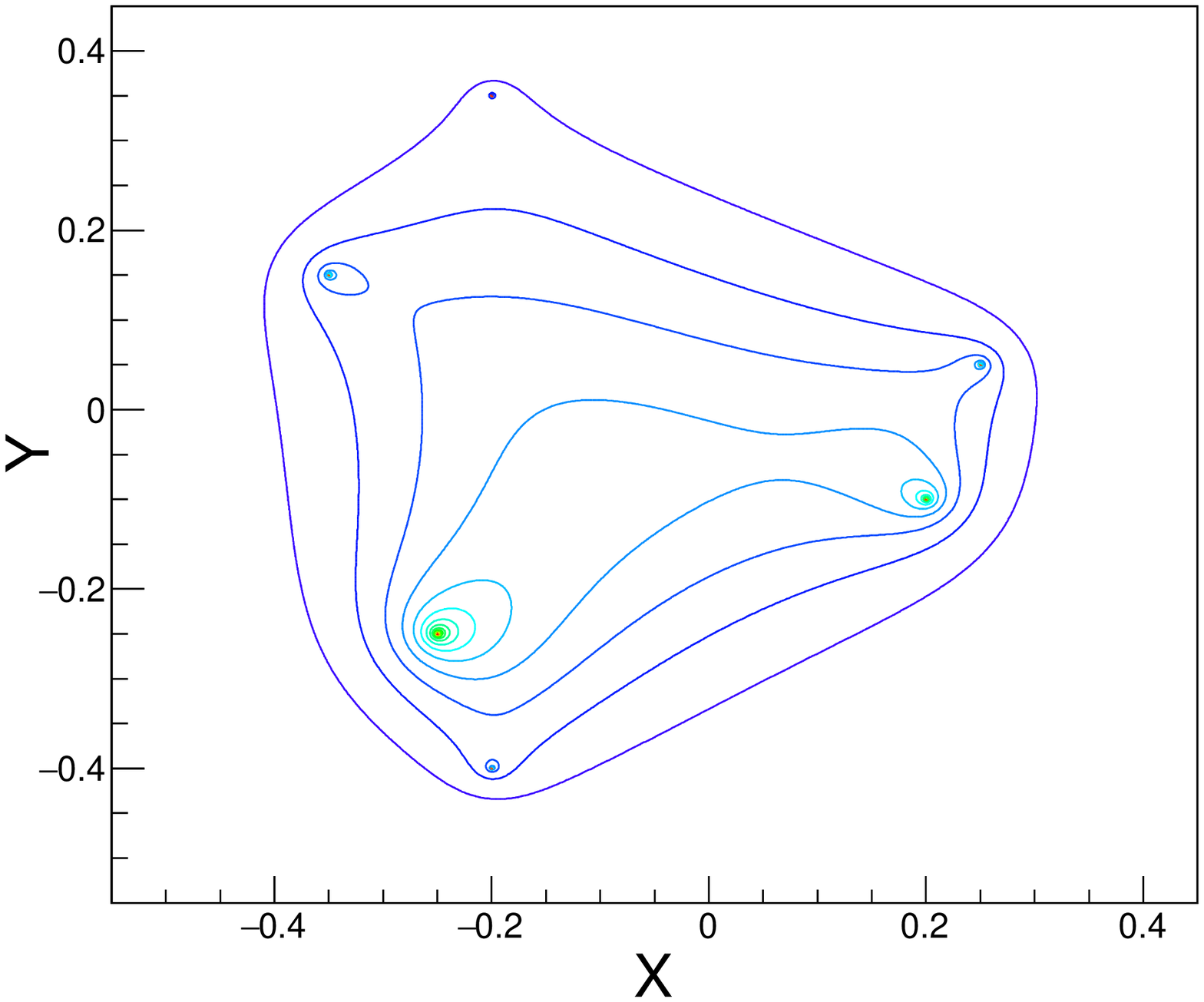}	
\caption{(Color online) Gauge fixed slices for the quarks
          at $(-0.2, 0.35, 0.0)$, $(-0.35, 0.15, 0.0)$, $(-0.25, -0.25, 0.0)$,
          $(-0.2, -0.4, 0.0)$, $( 0.2, -0.1, 0.0)$, and $(0.25, 0.05, 0.0)$.
}
 \label{fig1}
 \end{figure}

For topologically non-trivial vacuum states, the strong interaction Lagrangian is given by
\begin{equation}
 \mathcal{L} = -\frac{1}{4}F_{\mu\nu}^{\alpha}F_{\alpha}^{\mu\nu}
               -\frac{\theta g^{2}}{32\pi^{2}}F_{\mu\nu}^{\alpha}\tilde{F}_{\alpha}^{\mu\nu}
               + \sum_{f} \bar{\psi_{f}}\left[i\gamma^{\mu}(\partial_{\mu}-igA_{\mu})-m_{f}\right]\psi_{f}.
\label{eq10}
\end{equation}
The $\theta$-term characterizes the vacuum state and the observations of parity-odd domains localized
in space and time strongly support the spacetime dependence of $\theta$ as $\theta=\theta(\mathbf{x},t)$.
In case of varying $\theta$ values, the vacuum domains can be classified by the conditions
$\theta(\mathbf{x},t) = \theta_{i}$ with fixed values of $\theta_{i}$.
The classified points constitute surfaces and according to the movements of quarks
the forms of surfaces will change. For example, the points with fixed $\theta_{i}$'s
for 6 quark configuration can be drawn as in Fig.~\ref{fig1}~\cite{R19}.
%
%
%
This configuration can be made in the scattering processes of proton-proton collisions.
Since the different surfaces correspond to different values of $\theta$,
we have to consider the strong interaction vacuum as formed by the union of different slices
in case of spacetime dependent $\theta$.
For a given slice, the instanton contribution can be estimated by
the value of dimension-2 condensate $\langle A_{\mu}^{2} \rangle$.
Then we can assign another value of $\langle A_{\mu}^{2} \rangle$ to another slice.
These configurations can be represented as
\begin{equation}
   \langle A_{\mu}^{2} \rangle_\theta = C_\theta
\label{eq11}
\end{equation}
with fixed $C_\theta$ for each slice. For different slices we can write
\begin{equation}
   \langle A_{\mu}^{2} \rangle = C(\mathbf{x},t),
\label{eq12}
\end{equation}
and the spacetime dependence of $\theta$ is transferred into that of $\langle A_{\mu}^{2} \rangle$.
The whole QCD vacuum is now composed of these slices establishing
the generalized $\theta$ vacuum~\cite{R20}.
%
%
%
\section{STRUCTURE OF GAUGE FIXED SLICES}
\label{sec3}
%
In general relativity, gauge invariance stems from the independence of the field equations
on the choice of coordinate system. The Einstein field equations depend only on metric tensor
and energy-momentum of the system, and therefore the choice of a coordinate system does not
affect the equations of motion.
However, in order to solve the equations of motion it is quite important to choose
the coordinate system and this choice of particular system corresponds to
the fixing of particular gauge~\cite{R21}.
For quantum field theories, the main variables are changed into the field variables
at each space-time point, and since the field variables generally include
internal degrees of freedom the gauge fixing condition becomes functional relation
specifying the surface on which each gauge orbit has unique position.
The gauge orbit is obtained by connecting the field variables that can be related
by gauge transformations. Thus the determination of functional relation for gauge
fixing is a dual process with respect to the following up the gauge orbit~\cite{R22}.

For the space of gluonic field $\mathbf{A}_{\mu}$,
the condition for gauge fixing can be written as
\begin{equation}
   f(\mathbf{A}_{\mu}) = 0.
\label{eq13}
\end{equation}
Traditionally the functional form of $f$ had been taken to be linear in
$\mathbf{A}_{\mu}$ because of simplicity and the corresponding surface is
taken to be a plane in the field space.
However, now we have the generalized $\theta$ vacuum defined by the union of
gauge slices given in Eq.(\ref{eq12}) where the functional form is quadratic.
The quadratic form originates from the existence of dimension-2 condensate due to
the nonperturbative self-interactions of gluons.
These gluonic self-interactions induce that the gluonic states cannot be
classified according to the number of gluons and the unique possibility to
define the gluonic state is to consider the state of infinite number of gluons.
Let this state be represented by $ | \Omega \rangle$ and then we have
\begin{equation}
   A_{\mu}^{a}(x) | \Omega \rangle = \beta | \Omega \rangle
\label{eq14}
\end{equation}
because the gluon number is not changed by one creation or by one annihilation.
We get the covariant relation
\begin{equation}
   \langle \Omega | A_{\mu}^{a}(x) A_{a}^{\mu}(x)| \Omega \rangle = \alpha^2
\label{eq15}
\end{equation}
with inner producting each side. This condition is equivalent to the gauge slice condition
in Eq.(\ref{eq12}) if we take the state $ | \Omega \rangle$ as one vacuum state.
For different $\alpha$ values we have independent vacuum states represented by
different values of dimension-2 condensate.
In ordinary flat space, these vacuum states might be overlapped,
however, because the vacuum condensates have to be considered
only around hadronic structure as indicated by in-hadron condensates,
the QCD vacuum is taken to be the union of curved slices fixed
by the value of dimension-2 condensate.

In order to figure out the structure of gauge fixed slices, we need to classify
the gluonic domains formed by slices with given values of dimension-2 condensate.
Because the gluonic domains are made around quarks and antiquarks which behave as
sources and sinks of the gluons, we can classify the domains by the numbers of
quarks and antiquarks in each domain. If we represent the gluonic domain
around $a$ quarks and $b$ antiquarks as $R_{a, \bar{b}}$,
then meson domain is represented by $R_{1, \bar{1}}$, and
baryon domain corresponds to $R_{3, \bar{0}}$.
In this representation we can include arbitrary number of quarks and antiquarks
such as $R_{2, \bar{3}}$, however, since the final states of strong interactions
are observed as hadrons we need to categorize the gluonic domains
with definite rules of combinations.
Two gluonic domains are united when one quark-antiquark pair annihilates
and conversely one gluonic domain can be divided into two domains
by one quark-antiquark pair creation.
Therefore the union and the intersection operations can be defined into
the sets of gluonic domains, which can be formulated as follows~\cite{R23}:
\newcommand{\squishlist}{
 \begin{list}{$\bullet$}
  { \setlength{\itemsep}{0pt}
     \setlength{\parsep}{0pt}
     \setlength{\topsep}{2pt}
     \setlength{\partopsep}{0pt}
     \setlength{\leftmargin}{1.5em}
     \setlength{\rightmargin}{1.5em}
     \setlength{\labelwidth}{1em}
     \setlength{\labelsep}{0.5em} } }
\newcommand{\squishend}{
  \end{list}  }
\squishlist
\item Gluonic domains are open sets.
\item The union of gluonic domains is a glunoic domain.
\item The intersection between a connected gluonic domain and
      disconnected gluonic domains is the reverse operation of the union.
\squishend
These are just the conditions needed to construct the topological spaces of gluonic domains.
The constructed topological space encompassing $i$ baryons and $j$ antibaryons
can be represented as
\begin{equation}
T_{i,\bar{j}}~ = ~\{ \emptyset, ~R_{3,\bar{0}}^{i}R_{0,\bar{3}}^{j},
   ~R_{3,\bar{0}}^{i-1}R_{0,\bar{3}}^{j-1}R_{2,\bar{2}},
   ~R_{3,\bar{0}}^{i-2}R_{0,\bar{3}}^{j-2}R_{2,\bar{2}}^{2}, \cdots \}.
\label{eq16}
\end{equation}

\begin{figure}
 \includegraphics[width=0.76\linewidth]{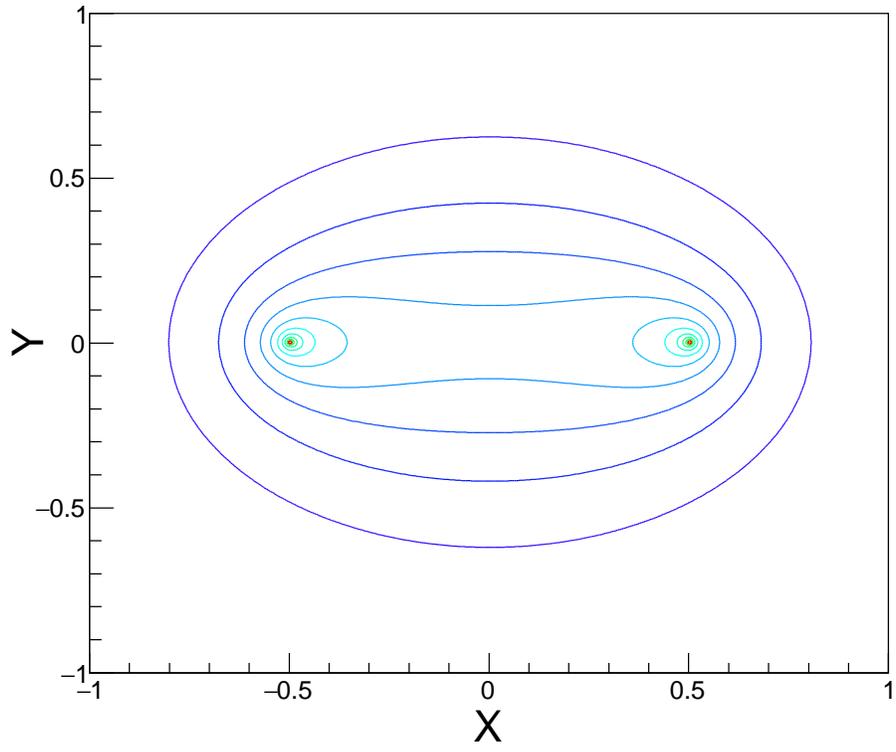}
  \caption{(Color online)
      Gauge slices for the quarks at $(-0.5, 0.0, 0.0)$ and $(0.5, 0.0, 0.0)$
          with $\beta = 1.0$ and $k = 1.0$.}	
 \label{fig2}
 \end{figure}

For the constructed topological spaces, we may introduce a systematic measure
to deduce the forms of the gauge fixed slices defined by the values of
dimension-2 condensate. The measure is related to the nature of QCD vacuum
and so it has to be represented by quantities characterizing the vacuum.
Because the characteristics of vacuum states have been expressed as various condensates,
it is plausible to compare two fundamental condensates to deduce the form of
gauge fixed slice. The two fundamental condensates can be chosen
as the quark condensate and the dimension-2 gluon condensate.
The nonlocal quark condensate is defined by~\cite{R24}
\begin{equation}
\langle : \bar{q}(x)U(x,0)q(0) : \rangle \equiv \langle : \bar{q}(0)q(0) : \rangle Q(x^{2}),
\label{eq17}
\end{equation}
where $U(x,0)$ represents the connection through the gluonic domain.
In the gluonic domain, the gluonic effects are parametrized by the value of
$\langle A^2_\mu \rangle$ and this value changes from point to point.
The comparison with quark condensate can be made possible by assuming
\begin{equation}
 \langle : \bar{q}(x)U(x,y)A^{a}_{\mu}(y)A^{\mu}_{a}(y)U(y,0)q(0) : \rangle \propto
 \langle : \bar{q}(x)U(x,y)q(y)\bar{q}(y)U(y,0)q(0) : \rangle ,
\label{eq18}
\end{equation}
where the final result can be expressed as products of two $Q$'s.
To derive the functional form of $Q(x^2)$ we consider a measure $\mathfrak{M}(Q)$.
In general, we can assume that the value of $Q$ decreases as the distance $x$
between the quarks increases. This assumption can be stated as
\begin{equation}
  \mathfrak{M}(Q)~~\text{increases ~as}~~Q~~\text{decreases.}
\label{eq19}
\end{equation}
Another condition can be written down by comparing both sides of Eq.(\ref{eq18})
resulting in
\begin{equation}
  \mathfrak{M}(Q_1Q_2) = \mathfrak{M}(Q_1) + \mathfrak{M}(Q_2).
\label{eq20}
\end{equation}
Then these two conditions lead to the solution
\begin{equation}
  \mathfrak{M}(Q) = -k \ln\frac{Q}{Q_0},
\label{eq21}
\end{equation}
where $k$ is an appropriate parameter and $Q_0$ is a normalization constant.
When we try to represent the measure $\mathfrak{M}(Q)$ as a metric function,
we may use the distance function
\begin{equation}
    d(\mathbf{x},\mathbf{y})= \left|\mathbf{x}-\mathbf{y} \right|^{\nu}
\label{eq22}
\end{equation}
with $\mathbf{x}$ and $\mathbf{y}$ representing the positions of the quark pair.
We need to sum over the contributions from different values of $\nu$
and the full amplitude becomes
\begin{equation}
  Q = {Q_0}\exp\left\{-\frac{1}{k} \int_{1}^{\alpha}F(\nu)r^{\nu} d\nu \right\}.
\label{eq23}
\end{equation}
The weight factor $F(\nu)$ has been introduced to account for
possible different contributions from different $\nu$'s, and the variable $r$ is
\begin{equation}
    r = \frac{1}{\ell}\left|\mathbf{x} - \mathbf{y} \right|
\label{eq24}
\end{equation}
with $\ell$ being a scale parameter.
For the case of equal weight $F(\nu) = 1 $ and for $\alpha = 2$, we get
\begin{equation}
  Q = Q_{0} \exp \left\{ -\frac{1}{k} \frac{r^{2} - r }{\ln r} \right\}.
\label{eq25}
\end{equation}
Perturbative local effects can be included by substituting $r^{\beta}Q$ for $Q$,
and then $Q$ becomes
\begin{equation}
 Q = \frac{Q_0}{r^\beta} \exp \left\{ -\frac{1}{k} \frac{r^2 - r}{\ln r} \right\}.
\label{eq26}
\end{equation}
Introducing the notation
\begin{equation}
 Q_{ij} = \frac{Q(r_{ij})}{Q_0} = \frac{Q(\left|\mathbf{r}_{i}-\mathbf{r}_{j} \right|)}{Q_0},
\label{eq27}
\end{equation}
we can write the value of $\langle A^2_\mu \rangle$ at the point $\mathbf{x}$ as
\begin{equation}
 \langle A_{\mu}^{2} \rangle = A_{0}^{2} \prod_{i=1}^{6} Q_{xi} \left\{
                                         \sum_{i=1}^{6} {\prod_{r_j , r_k \ne r_i}} Q_{jk}
                                        + \sum_{r_j ,r_k}Q_{jk} \cdot {\prod_{r_\alpha , r_\gamma \ne r_j , r_k}}
                                        Q_{\alpha\gamma} \right\}
\label{eq28}
\end{equation}
for the case of a 6-quark domain and several slices are shown in Fig.~\ref{fig1}.
For a meson domain $R_{1,\bar{1}}$, curved slices are shown in Fig.~\ref{fig2}.
%
%
%
\section{GLUON PROPAGATOR ON CURVED GAUGE SLICE}
\label{sec4}
%
%
Before we discuss the gluon propagation in curved space,
it is valuable to summarize the changes of ordinary gluon propagator
when the dimension-2 condensate $\langle A^2_\mu \rangle$ exists.
For simplicity we assume that the space is flat and the value of
$\langle A^2_\mu \rangle$ is taken to be a constant.
Because the condensate value is taken to be constant,
we may decompose the gluon field as
\begin{equation}
    A_{\mu}^{a}(x)= \mathbf{A}_{\mu}^{a} + \mathcal{A}_{\mu}^{a}(x),
\label{eq29}
\end{equation}
where $\mathbf{A}_{\mu}^{a}$ contributes to the condensate as the zero-momentum mode.
Then, the effective Lagrangian density for gluons can be written as~\cite{R25}
\begin{equation}
 \mathcal{L}_{G} = -\frac{1}{4}\mathcal{G}_{\mu\nu}^{a}(x)\mathcal{G}_{a}^{\mu\nu}(x)
                   +\frac{m_{G}^{2}}{2}\mathcal{A}_{\mu}^{b}(x)\mathcal{A}_{b}^{\mu}(x)
\label{eq30}
\end{equation}
with
\begin{equation}
    \mathcal{G}_{\mu\nu}^{a}(x) = \partial_{\mu}\mathcal{A}_{\nu}^{a}(x)
    -\partial_{\nu}\mathcal{A}_{\mu}^{a}(x)+gf^{abc}\mathcal{A}_{\mu}^{b}(x)\mathcal{A}_{\nu}^{c}(x).
\label{eq31}
\end{equation}
The gluon mass $m_G$ is defined by the condensate value
\begin{equation}
 \left\langle \text{vac} \left|  \mathbf{A}^{a}_{\mu} \mathbf{A}^{b}_{\nu} \right| \text{vac} \right\rangle
  = -g_{\mu\nu}\delta_{ab}\frac{m_{G}^{2}}{9g^2} ,
\label{eq32}
\end{equation}
and the existence of mass term indicates that the polarization tensor for a gluon has the contribution
\begin{equation}
 \Pi_{\mu\nu(\text{mass})}^{ab}
    = \delta_{ab}\left(g_{\mu\nu} - \frac{k_{\mu}k_{\nu}}{k^2}\right)m_{G}^2.
\label{eq33}
\end{equation}
If we define the scalar function $\Pi(k^2)$ by
\begin{equation}
 \Pi_{ab}^{\mu\nu}(k) = \left(g^{\mu\nu}k^2 - k^{\mu}k^{\nu}\right)\Pi(k^2)\delta_{ab},
\label{eq34}
\end{equation}
then $\Pi(k^2)$ can be written as
\begin{equation}
 \Pi(k^2) = \frac{m_{G}^2}{k^2} + \frac{\Pi_{2}(k^2)}{k^2} .
\label{eq35}
\end{equation}
The first term represents the mass effects originating from the vacuum condensate
corresponding to the Schwinger mechanism, and the second term appears as the next corrections.
The form of $\Pi_{2}(k^2)$ is determined by the relation
\begin{equation}
 \left(g^{\mu\nu} - \frac{k^{\mu}k^{\nu}}{k^{2}}\right)\left[ \frac{\Pi_{2}(k^2)}{1 - \frac{\Pi_{2}(k^2)}{k^{2}-m_{G}^2} } \right]\delta_{ab}
 = i \int d^{4}x~\mbox{e}^{ik \cdot x} \left\langle \text{vac} \left| T \left[\hat{J}^{\mu}_{T, a}(x)\hat{J}^{\nu}_{T, b}(0)\right] \right| \text{vac} \right\rangle ,
\label{eq36}
\end{equation}
where
\begin{equation}
    \hat{J}_{T,a}^{\mu}(x) \equiv \left(g^{\mu\nu} - \frac{k^{\mu}k^{\nu}}{k^{2}}\right)J_{\nu,a}(x)
\label{eq37}
\end{equation}
with
%
\begin{eqnarray}
 J_{\nu}^{a}(x) = gf^{abc}\left[ A_{b}^{\mu}(x)\partial_{\nu}A_{\mu}^{c}(x)
                                - 2A_{b}^{\mu}(x)\partial_{\mu}A_{\nu}^{c}(x)
                                - A_{\nu}^{c}(x)\partial^{\mu}A_{\mu}^{b}(x) \right]  \nonumber \\
                 + ~~g^{2}f^{abc}f^{a^\prime b^\prime c}\left[A_{\mu}^{b}(x)A_{b^\prime}^{\mu}(x)A_{\nu}^{a^\prime}(x)\right].
\label{eq38}
\end{eqnarray}
%
%
By putting the condition in  Eq.(\ref{eq29})
and using the relation in  Eq.(\ref{eq32}), we obtain
\begin{equation}
 i \int d^{4}x~\mbox{e}^{ik \cdot x} \left\langle \text{vac} \left| T \left[\hat{J}^{\mu}_{T, a}(x)\hat{J}^{\nu}_{T, b}(0)\right] \right| \text{vac} \right\rangle
   = \left( g^{\mu\nu} - \frac{k^{\mu}k^{\nu}}{k^2} \right)\delta_{ab}
     \left[ \frac{-\frac{4}{3}m_{G}^{2}k^2}{k^2 - m_{G}^{2} - \Pi_{2}(k^2)} \right],
\label{eq39}
\end{equation}
and then we get
\begin{equation}
 \Pi_{2}(k^2) = - \frac{\frac{4}{3}m_{G}^2 k^2}{k^2 - m_{G}^2} .
\label{eq40 }
\end{equation}
With these results we can write the form of transverse gluon propagator as
\begin{equation}
  D_{T, ab}^{\mu\nu}(k)
   = \left( g^{\mu\nu} - \frac{k^{\mu}k^{\nu}}{k^2} \right)\delta_{ab}
     \frac{1}{k^2 - m_{G}^{2} +\frac{4}{3}\frac{m_{G}^{2}k^2}{k^2 - m_{G}^2} } ,
\label{eq41}
\end{equation}
where we can see that there exists no pole and so the gluons cannot propagate freely
reaching far away from the source. This damping of gluon propagator originates
from the constant dimension-2 condensate generating dynamical mass of gluon.
Phenomenologically the dynamical gluon mass is determined to be
about $500 \sim 600$ MeV~\cite{R26},
however, these values are too large to be considered as the value of just one gluon
because there exist so many gluons in a hadron.
Moreover, the gluonic structure of hadron is not uniform as assumed in Eq.(\ref{eq32}),
so that we need to quantize the gluonic field on the curved gauge slices
as shown in Fig.~\ref{fig1} and in Fig.~\ref{fig2}.

The gauge slices are defined by the Eq.(\ref{eq15}) with different $\alpha$ values
and they turn out to be curved surfaces around quarks and antiquarks forming hadronic structure.
In order to account for the curvature we need to replace the ordinary derivative
by the covariant derivative.
The covariant derivative is defined as
\begin{equation}
    D_{\mu}\mathbf{A}_{\nu}
        \equiv \frac{\partial \mathbf{A}_{\nu}}{\partial x^{\mu}} - \Gamma_{\nu\mu}^{\kappa}\mathbf{A}_\kappa  ,
\label{eq42}
\end{equation}
where $\Gamma_{\nu\mu}^{\kappa}$ represents the affine connection which relates the ordinary coordinates
to the freely falling coordinates. When $\Gamma_{\mu\nu}^{\kappa} = \Gamma_{\nu\mu}^{\kappa}$,
the difference between two covariant derivatives is given by
\begin{equation}
    D_{\mu}\mathbf{A}_{\nu} - D_{\nu}\mathbf{A}_{\mu}
     =  \frac{\partial \mathbf{A}_{\nu}}{\partial x^{\mu}} - \frac{\partial \mathbf{A}_{\mu}}{\partial x^{\nu}},
\label{eq43}
\end{equation}
which implies that covariant curl is the same as ordinary curl and there appears
no curvature effect in first order derivatives. However, the Lagrangian which is used
in the derivation of the gluon propagator contains the second order derivatives and
the difference between two second order derivatives becomes~\cite{R27}
\begin{equation}
    D_{\kappa}D_{\nu}\mathbf{A}_{\mu} - D_{\nu}D_{\kappa}\mathbf{A}_{\mu}
     = -\mathbf{A}_{\sigma}R_{~\mu\nu\kappa}^{\sigma},
\label{eq44}
\end{equation}
where the Riemann-Christoffel curvature tensor $R_{~\mu\nu\kappa}^{\sigma}$ is defined as
\begin{equation}
    R_{~\mu\nu\kappa}^{\sigma} \equiv \frac{\partial \Gamma_{\mu\nu}^{\sigma}}{\partial x^\kappa}
                                 -\frac{\partial \Gamma_{\mu\kappa}^{\sigma}}{\partial x^{\nu}}
                                 + \Gamma_{\mu\nu}^{\eta}\Gamma_{\kappa\eta}^{\sigma}
                                 - \Gamma_{\mu\kappa}^{\eta}\Gamma_{\nu\eta}^{\sigma}.
\label{eq45}
\end{equation}
The Ricci tensor is given by
\begin{equation}
  R_{\mu\kappa} = R_{~\mu\sigma\kappa}^{\sigma},
\label{eq46}
\end{equation}
and the curvature scalar is defined as $R = g^{\mu\nu}R_{\mu\nu}$.

Now the free Lagrangian for gluon propagator in curved space can be written as
\begin{equation}
 L_D  = -\frac{1}{4} \int d^{4}x~ \left(D_{\mu}\mathbf{A}_{\nu} - D_{\nu}\mathbf{A}_{\mu}\right)
                            \cdot \left(D^{\mu}\mathbf{A}^{\nu} - D^{\nu}\mathbf{A}^{\mu}\right).
\label{eq47}
\end{equation}
In order to carry out the functional integration we have to arrange
the differential operators into the form
\begin{equation}
    L_D = \frac{1}{2} \int d^{4}x~ \mathbf{A}_{\mu}
           \cdot \left(D^2 g^{\mu\nu} - D^\nu D^\mu \right) \mathbf{A}_{\nu},
\label{eq48}
\end{equation}
where the covariant differential operators do not commute as in Eq.(\ref{eq44}).
To compare with ordinary gluon propagator in flat space we rearrange
the differential operators and get
\begin{equation}
   L_D = \frac{1}{2} \int d^{4}x~ \mathbf{A}_{\mu}
           \cdot \left(D^2 g^{\mu\nu} - D^\mu D^\nu + R^{\mu\nu} \right) \mathbf{A}_{\nu}
\label{eq49}
\end{equation}
with additional Ricci tensor $R^{\mu\nu}$.
The gluon propagator is obtained as the inverse matrix of the tensor
that appears between the two gluonic fields.
In flat space, the differential operators can be replaced by momentum variables
and the inverse matrix can be calculated easily.
However, in curved space, the appearance of Ricci tensor makes it non-trivial
to calculate the inverse matrix and we need to introduce additional assumptions
to deduce the form of gluon propagator.
The simplest case is that of flat space in which the Ricci tensor vanishes
and we can get the ordinary form of gluon propagator.
The next case is that of maximally symmetric space~\cite{R28} where we can put
\begin{equation}
    R^{\mu\nu} = m^2 g^{\mu\nu}
\label{eq50}
\end{equation}
with appropriate parameter $m^2$. This parameter is definitely related to the
curvature scalar of the space and is obtained by considering free propagation of
massless gluon as a gauge boson. Then the Lagrangian becomes
\begin{equation}
   L_D = \frac{1}{2} \int d^{4}x~ \mathbf{A}_{\mu}
           \cdot \left[ (D^2 + m^2)g^{\mu\nu} - D^\mu D^\nu \right] \mathbf{A}_{\nu},
\label{eq51}
\end{equation}
and the gluon propagator can be obtained as
\begin{equation}
    D_{ab}^{\mu\nu}(k) = \frac{1}{k^2 - m^2} \left( g^{\mu\nu} - \frac{k^\mu k^\nu}{k^2} \right)\delta_{ab},
\label{eq52}
\end{equation}
which is just the propagator form of a particle with momentum $k$ and mass $m$.

In the usual process of dynamical mass generation of gluon propagator,
the polarization tensors are calculated by estimating various loop contributions
with appropriate renormalization constants.
Therefore the generated mass always has some uncertainty related to the renormalization process
and the loop expansions turn out to be meaningless for nonperturbative gluonic interactions.
Moreover the determined mass is too large to be considered as a correction to massless gluon.
Instead of these difficulties, it is quite natural to get the mass effects of gluons
from the curvature of gauge fixed slices.

%
%
\section{GLUONS AS DARK MATTER }
\label{sec5}
%

The curved gauge slices can be drawn not only for hadrons such as mesons and baryons
but also for any strongly interacting systems of quarks and gluons such as multiquark states
and nuclei and even for the system of quark-gluon plasma.
Because gluons are taken to be propagating along these gauge fixed slices,
the curvature effects are essential in estimating the gluonic contributions
to strongly interacting systems.
For perturbative approximations, gluon propagators are assumed to be in the same form
as that of photons except for color factors, and therefore the main features in
asymptotically free region resembles those of Abelian gauge theory~\cite{R29}.
But in the infrared region, that is, in the confining region, nonperturbative effects
play important roles in changing the vacuum states with nonvanishing condensate values.
These changed vacuum states can be classified according to the configuration of the source quarks.

For stable hadrons, the configurations of quarks are very simple.
In a meson, only one quark-antiquark pair exists and we can easily draw the gauge slices
as in Fig.~\ref{fig2}. However, when we want to probe the internal structure of the meson,
the main scattering processes are related with the quarks and the asymptotically free gluons
near the quarks leaving aside the whole bunch of gluons described as gauge slices.
If the probing particle has the enough energy to approach the asymptotically free region,
it can create quark-antiquark pairs and the created quarks and antiquarks change the
structure of gauge slices and form the final states with new gauge slices.
Because the final states are counted by hadrons instead of quarks and gluons,
the nonperturbative gluons described as gauge slices behave just like the dark matter.
They were not probed but contributed to the total mass of the final states.

In a baryon, three quarks have been introduced to account for the quantum numbers
but their masses could not be determined in a unique way.
The simplest way to determine the quark mass is to divide the total mass of a baryon
by 3 and the determined mass is named as constituent quark mass~\cite{R30}.
Although the lowest baryon states, the proton and the neutron,
have nearly the same mass, the other excited baryon states have quite different masses
and therefore the value of constituent quark mass has to be different from baryon to baryon.
These difficulties have lead to the idea of dynamical quark mass~\cite{R31}
including the effects of gluonic contributions, however,
in this case also, the probed physical mass is assigned
to the quark leaving the gluons as dark matter.
This situation is not changed in scattering processes.
In scattering processes, baryons are treated as composed of partons
which carry fractional momentum of the baryon and the first approximations
were made in such a way that the quarks were taken to be the partons without
considering the roles of gluons. These approximations resulted in various forms
of quark distribution functions~\cite{R32} but could not correspond to the experimental data.
The corrections to the first approximations had been made in the direction of
introducing sea quarks and there appeared various models with different forms of
valence quark and sea quark distribution functions~\cite{R33}.
Because the probing particles for the internal structures of baryons were leptons
such as electrons or neutrinos, the main processes occurred through the exchange
of photons or weak gauge bosons. Since the gluons did not interact directly
with these gauge bosons, it was natural to introduce sea quarks instead of gluons to
account for the final states.
However, the total momentum carried by charged partons turned out to be only
one half of the total momentum of the probed baryon.
The neutral partons had to be introduced playing the role of dark matter.

The hadronic masses are basically measured by the trajectories in magnetic fields
and then compared with other scattering data. The trajectories in magnetic fields
are determined by the charge and the momentum of the hadron.
The charge of a hadron is taken to be the sum of charges carried by valence quarks,
and the momentum of the hadron is measured by the velocity of the center of mass of the
hadron and the total rest mass of the hadron.
But the total rest mass of a hadron is not given by the sum of rest masses of the
valence quarks, and in fact, consists mostly of the intrinsically massless gluons
moving around. In order to deduce the rest masses of valence quarks,
we need to get rid of strong interactions and by considering electromagnetic and
weak interactions we can get the values
of 2.3~MeV for up quark and 4.8~MeV for down quark~\cite{R34}.
If we sum up these values, we have 9.4~MeV for a proton and 11.9~MeV for a neutron.
These values are just $1 \sim 1.3 \%$ of the total masses of proton and neutron.
Then from the viewpoint of measurement more than $98 \%$ of nucleon masses are dark matter
which is definitely composed of gluons.

To establish the standpoint of gluons as dark matter, we need to check whether
they can have survived from the Big Bang and can exist in galactic space
in a nonbaryonic state. It is helpful to use the arguments of maximum wavelength of
confined gluons applied to induce the idea of in-hadron condensates.
It has been suggested that gluons confined in a strongly interacting system
have a maximum wavelength of the order of the confinement scale~\cite{R35}
because they are confined within that system. Thus gluons confined in a hadron
have the maximum wavelength
\begin{equation}
    \lambda_{\text{max}} \simeq 1~\text{fm} \simeq \frac{1}{200~\text{MeV}},
\label{eq53}
\end{equation}
and those confined in a nucleus of diameter of 10~fm have
\begin{equation}
  \lambda_{\text{max}} \simeq 10~\text{fm} \simeq \frac{1}{20~\text{MeV}}.
\label{eq54}
\end{equation}
During the period of big-bang nucleosynthesis only light nuclei
such as $D$, $^{3}\text{He}$, $^{4}\text{He}$, and $^{7}\text{Li}$
have been formed, and so the maximum wavelength of the gluons confined
in such light elements can be estimated as
\begin{equation}
  \lambda_{\text{max}} \simeq 2~\text{fm} \simeq \frac{1}{100~\text{MeV}}.
\label{eq55}
\end{equation}
This implies that of the gluons created at the Big Bang only those with
energies greater than 100~MeV could be captured inside baryons or light nuclei.
Therefore the gluons with smaller energies had to be left as sterile gluons
if there existed no larger structure formed by strong interactions.
The heavier nuclei had been formed later in stars and in supernova explosions
and the gluons with energies between 20~MeV and 100~MeV could be captured
in those elements. Then how about the gluons with energies less than 20~MeV?
Are there any larger structures than the heavy nuclei interacting strongly
through exchanges of low energy gluons?

One possible candidate to accommodate the long wavelength gluons is the neutron star.
Neutron stars are known to be formed by gravitational collapse
after supernova explosions and the average density is approximately equal to the
density of nucleons. This means that the neutrons inside neutron star cannot exist
as independent neutrons but have to be in the form of overlapped configurations.
Thus the outer part of a neutron star is taken to be in the state of solids or liquids
composed of neutrons and the inner part in the state of quark-gluon plasma~\cite{R36}.
In the quark-gluon plasma region, the average distance between quarks is
smaller than that between quarks in neutrons.
Therefore the value of $\alpha^2$ in Eq.(\ref{eq15}) for the dimension-2 condensate
is expected to be larger in the inner part than in the outer part of the neutron star.
Because the neutron star is in the form of sphere, the gauge slices defined
by Eq.(\ref{eq15}) have to be spheres and the value $\alpha^2$ is expected to be
decreasing from the center of the neutron star.
In order to check out the decreasing pattern of the value of dimension-2 condensate,
we need to review the change of patterns for various configurations of quarks.

\begin{figure}
 \includegraphics[width=0.76\linewidth]{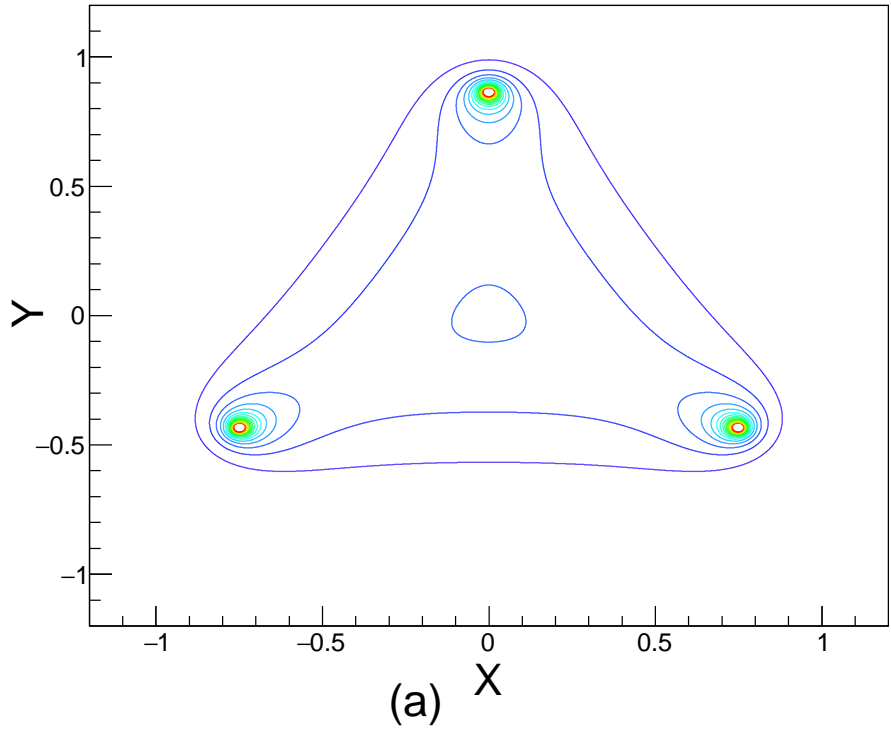}
 \includegraphics[width=0.76\linewidth]{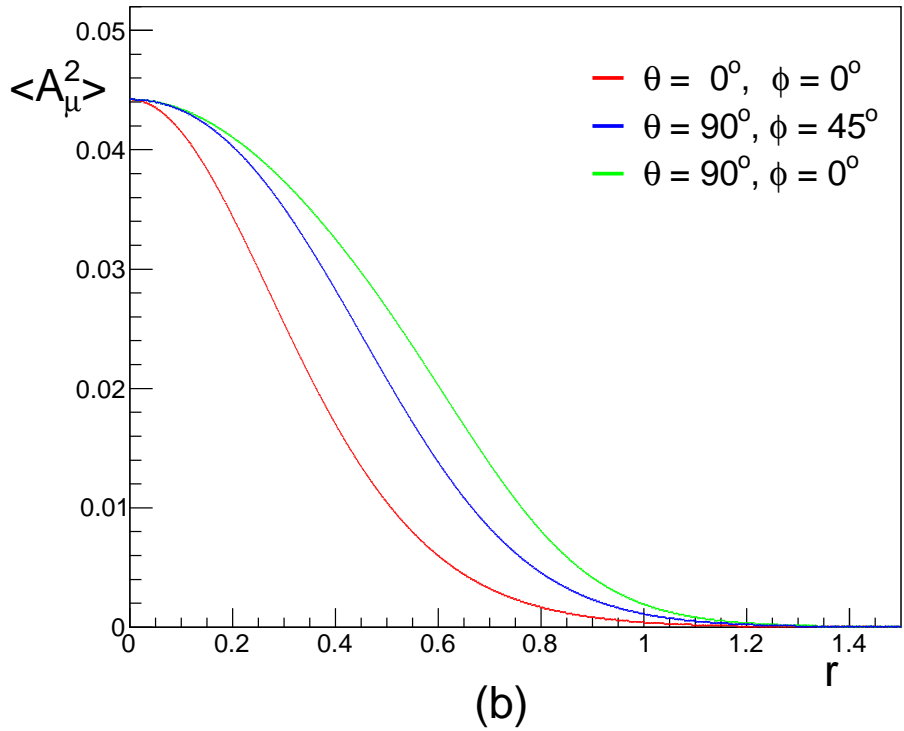}
 \caption{(Color online)
 (a) Equi-$\langle A_{\mu}^{2} \rangle$ surfaces for a baryon with quarks
   at $(0.0, \frac{\sqrt{3}}{2}, 0.0)$, $(-0.75, -\frac{\sqrt{3}}{4}, 0.0)$ and $(0.75, -\frac{\sqrt{3}}{4}, 0.0)$.
 (b) $\langle A_{\mu}^{2} \rangle$ values along the axes through the center of the triangle.
 }	
 \label{fig3}
 \end{figure}

\begin{figure}
 \includegraphics[width=0.49\linewidth]{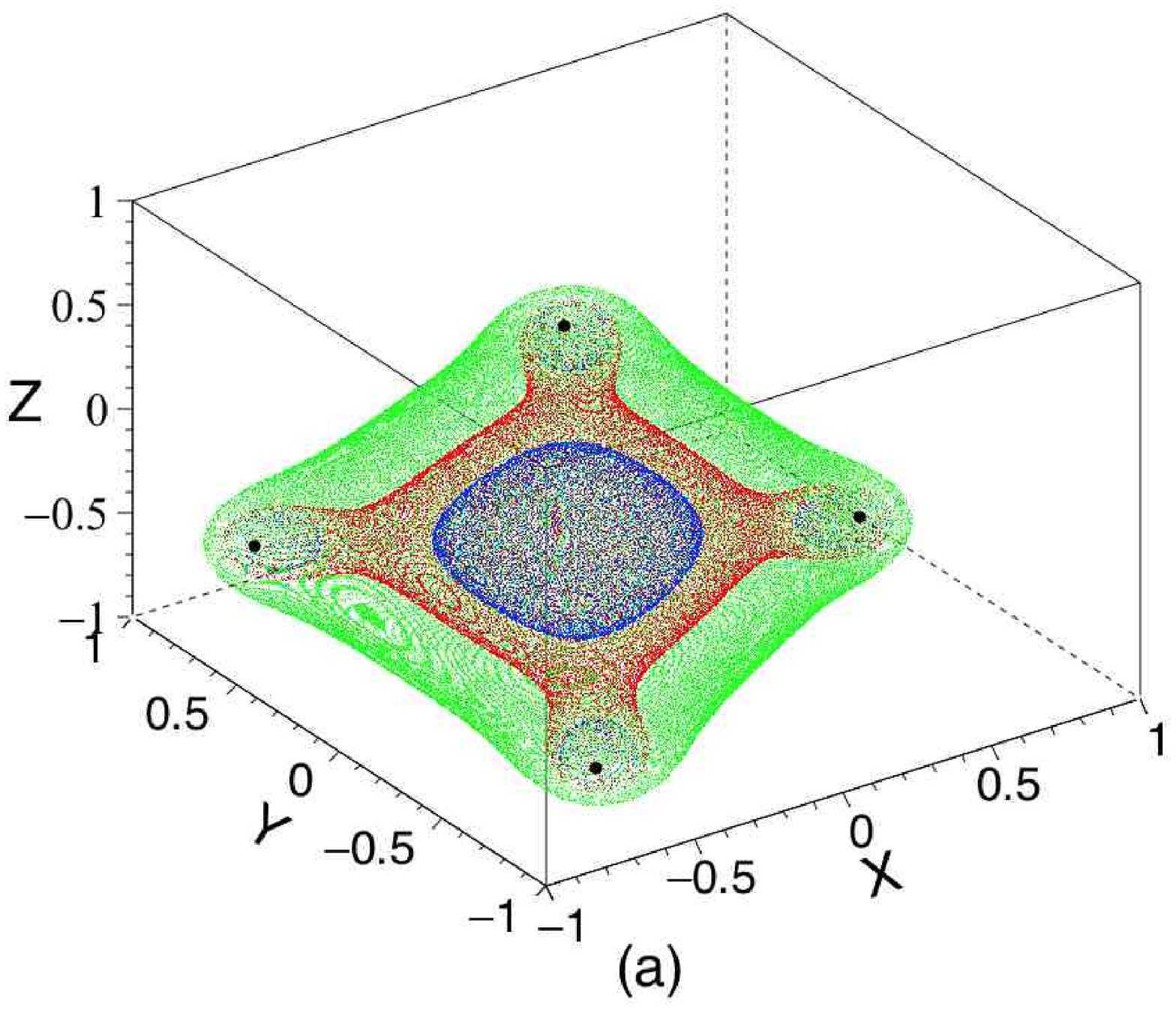}
 \includegraphics[width=0.498\linewidth]{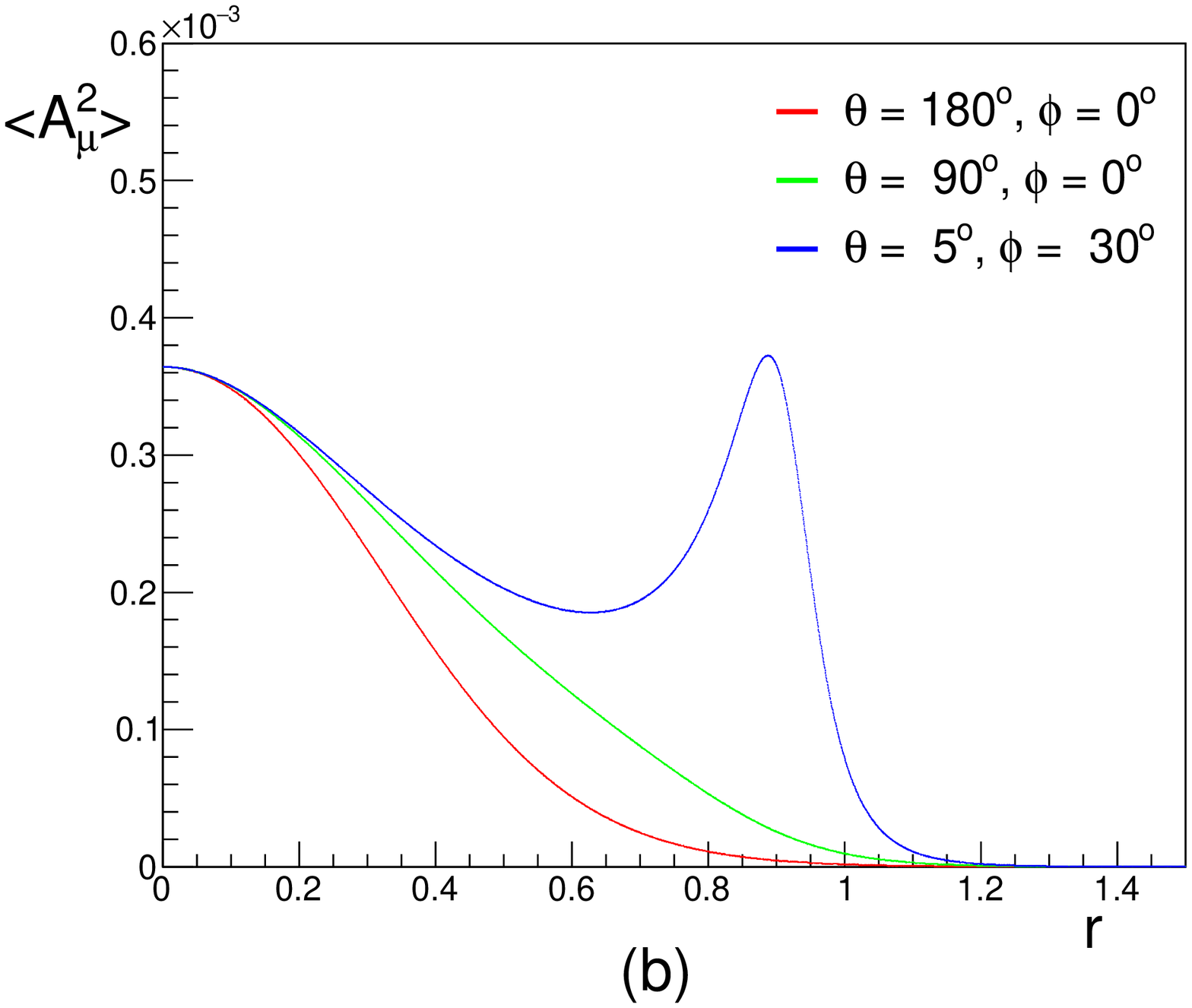}
 \caption{(Color online)
  (a) Equi-$\langle A_{\mu}^{2} \rangle$ surfaces for a tetraquark state with quarks
      at $(-\frac{\sqrt{3}}{4}, -0.75, -\frac{\sqrt{6}}{8})$, $(\frac{\sqrt{3}}{2}, 0.0, -\frac{\sqrt{6}}{8})$,
       $(-\frac{\sqrt{3}}{4}, 0.75, -\frac{\sqrt{6}}{8})$ and $(0.0, 0.0, \frac{3\sqrt{6}}{8})$.
  (b) $\langle A_{\mu}^{2} \rangle$ values along the axes through the center of the tetrahedron.
 }	
 \label{fig4}
 \end{figure}

The first example to check the variance of dimension-2 condensate is the baryon.
If we assume that the three quarks are located at the vertices of an equilateral triangle,
we can draw the equi-$\langle A_{\mu}^{2} \rangle$ surfaces as in Fig.~\ref{fig3}(a).
In Fig.~\ref{fig3}(b), the variations of $\langle A_{\mu}^{2} \rangle$ values
%
%
%
%
along the axes through the center of the triangle are
shown as a function of the distance from the center.
The next example is a tetraquark state which can be a candidate
for some $X$ particles~\cite{R37} discovered in the search of $(c\bar{c})$ excited states.
A typical tetraquark configuration is drawn in Fig.~\ref{fig4}(a)
and the variations of dimension-2 condensate are shown in Fig.~\ref{fig4}(b).
%
%
%
%

\begin{figure}
\includegraphics[width=0.49\linewidth]{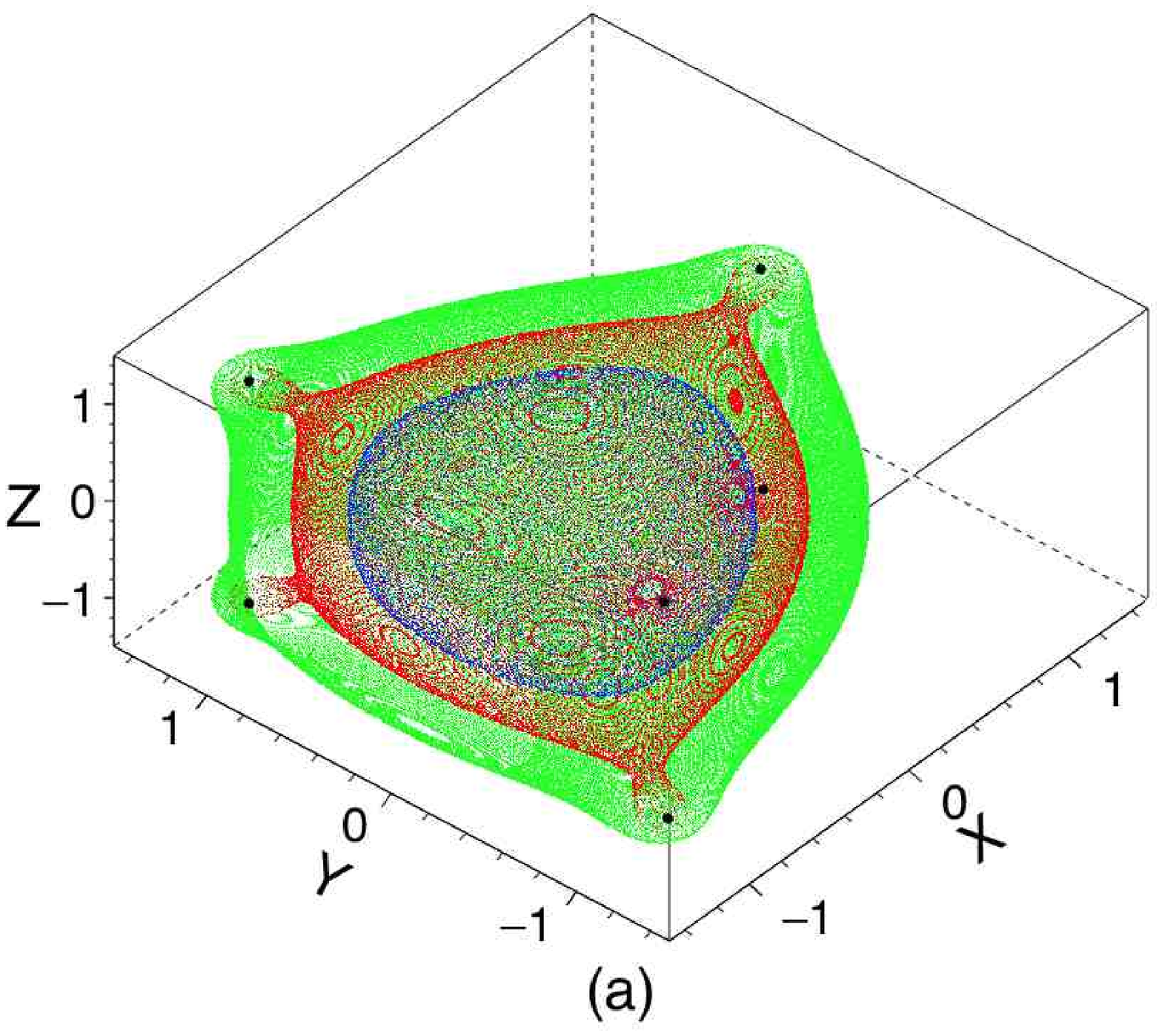}
\includegraphics[width=0.498\linewidth]{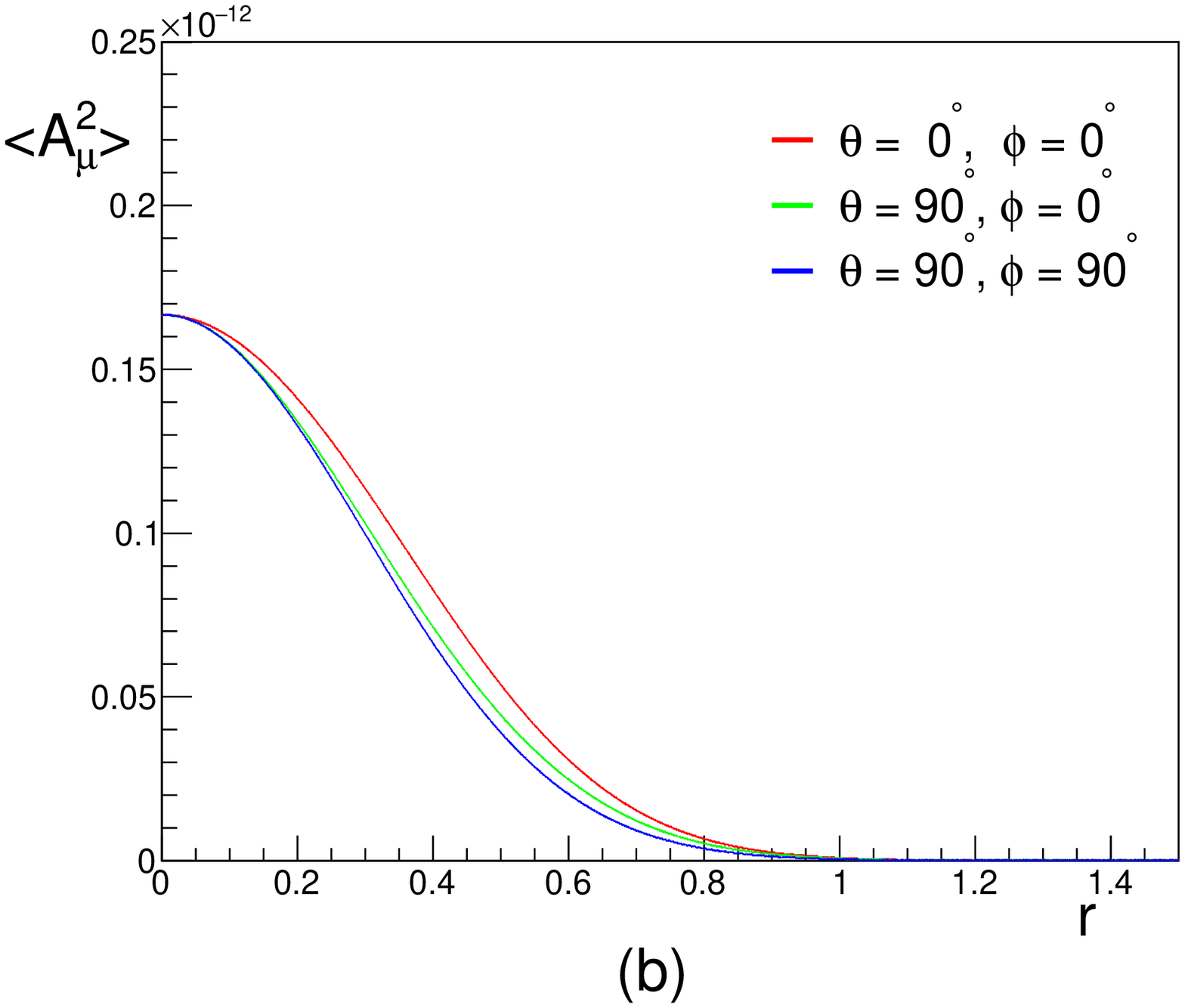}
\caption{(Color online)
  (a) Equi-$\langle A_{\mu}^{2} \rangle$ surfaces for a deuteron with quarks
      at $(-\frac{\sqrt{3}}{4}, -0.75, 0.75)$, $(\frac{\sqrt{3}}{2}, 0.0, 0.75)$,$(-\frac{\sqrt{3}}{4}, 0.75, 0.75)$,
      $(-\frac{\sqrt{3}}{4}, -0.75, -0.75)$, $(\frac{\sqrt{3}}{2}, 0.0, -0.75)$ and $(-\frac{\sqrt{3}}{4}, 0.75, -0.75)$.
  (b) $\langle A_{\mu}^{2} \rangle$ values along the $x-$axis, $z-$axis and the line with
      $\theta = 90^\circ$, $\phi = 90^\circ$.
}	
\label{fig5}
\end{figure}

\begin{figure}
 \includegraphics[width=0.49\linewidth]{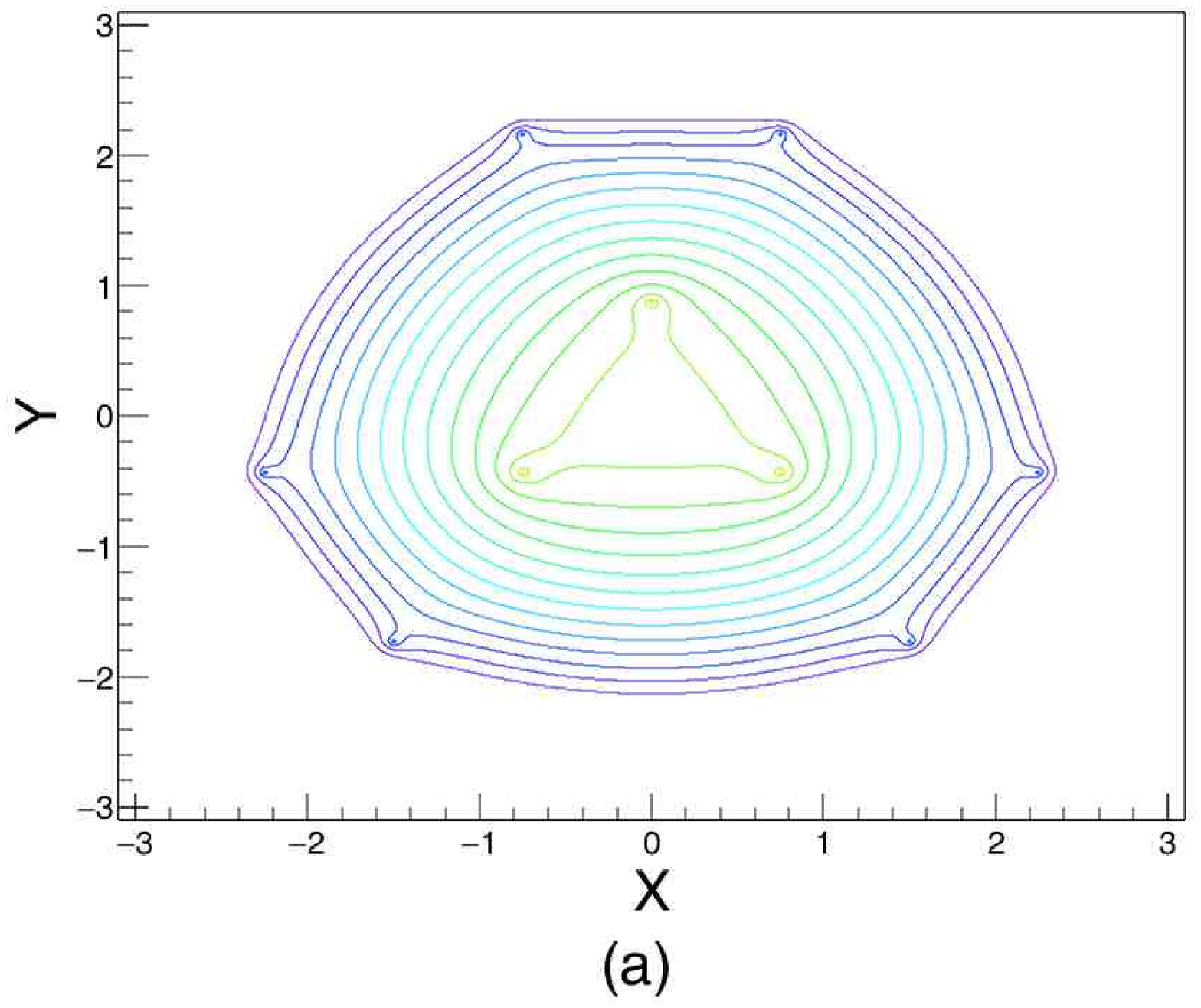}
 \includegraphics[width=0.498\linewidth]{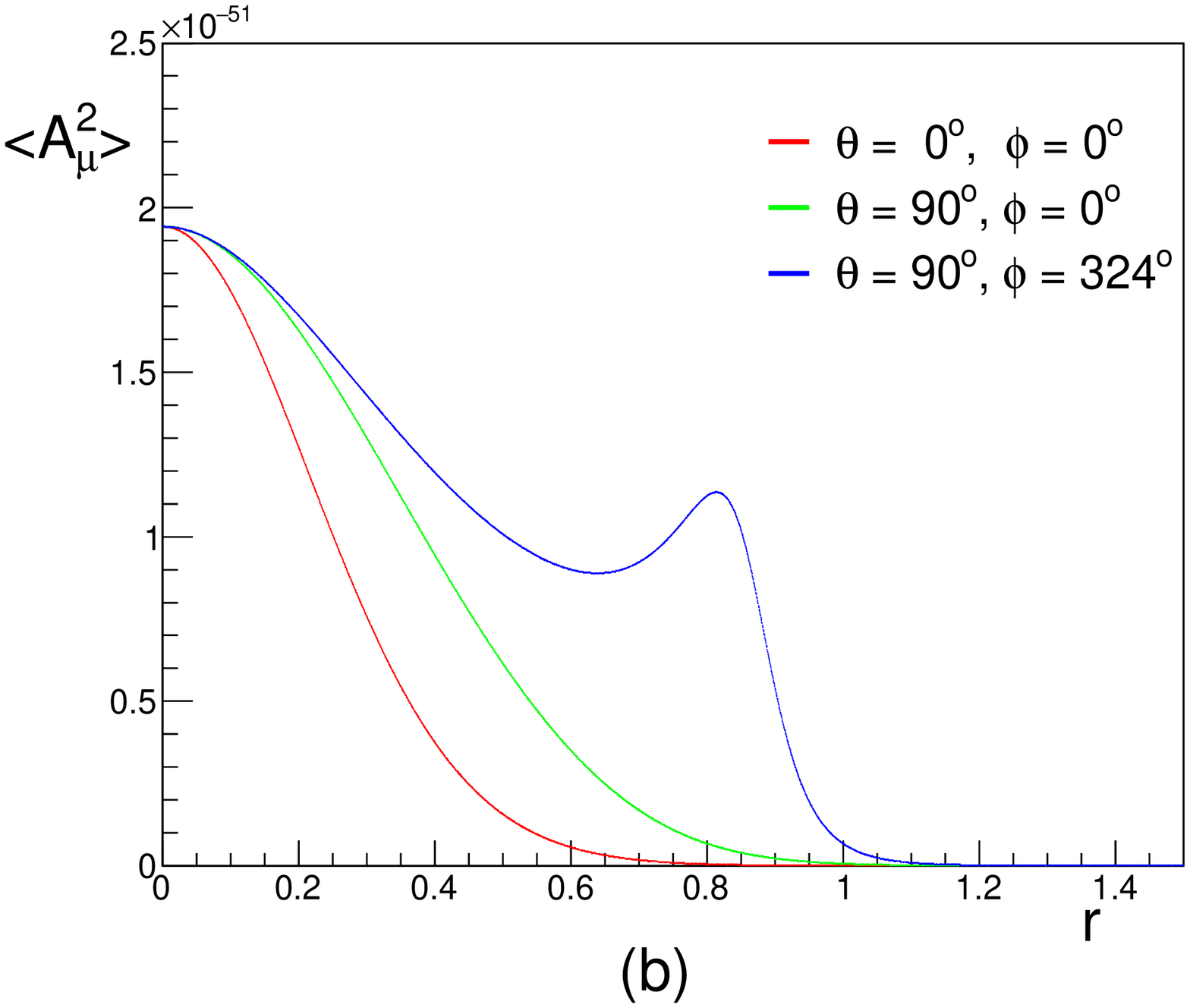}
   \caption{(Color online)
   (a) Equi-$\langle A_{\mu}^{2} \rangle$ surfaces for $^{3}\text{He}$ with quarks
      at $(0.0, \frac{\sqrt{3}}{2}, 0.0)$, $(-0.75, -\frac{\sqrt{3}}{4}, 0.0)$, $(0.75, -\frac{\sqrt{3}}{4}, 0.0)$,
      $(-0.75, \frac{5\sqrt{3}}{4}, 0.0)$, $(-2.25, -\frac{\sqrt{3}}{4}, 0.0)$, $(-1.50, -\sqrt{3}, 0.0)$,
      $(1.50, -\sqrt{3}, 0.0)$, $(2.25, -\frac{\sqrt{3}}{4}, 0.0)$ and $(0.75, \frac{5\sqrt{3}}{4}, 0.0)$.
   (b) $\langle A_{\mu}^{2} \rangle$ values along the $x-$axis, $z-$axis and the line with
      $\theta = 90^\circ$, $\phi = 324^\circ$.
  }	
 \label{fig6}
\end{figure}

\begin{figure}
 \includegraphics[width=0.49\linewidth]{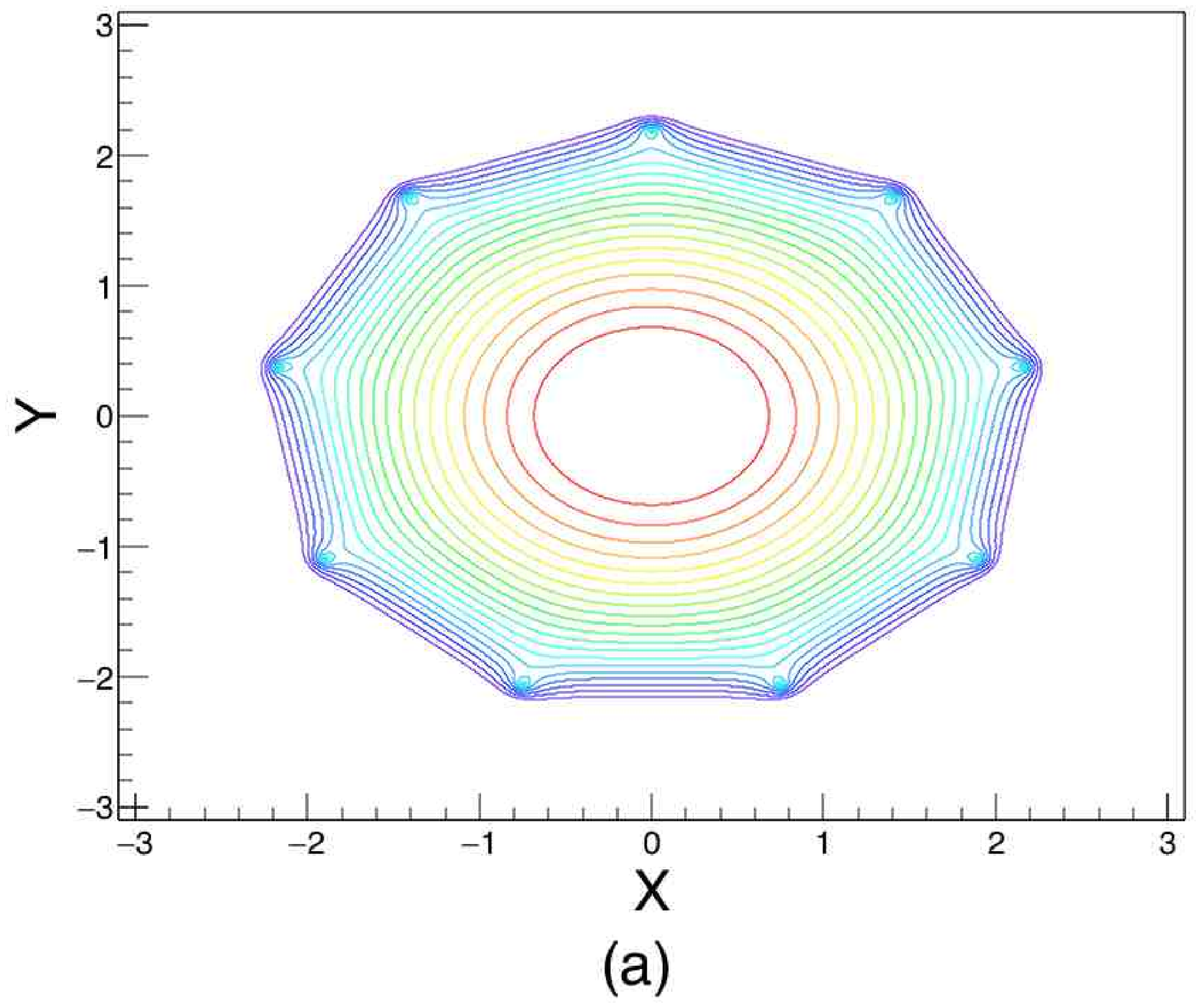}
 \includegraphics[width=0.498\linewidth]{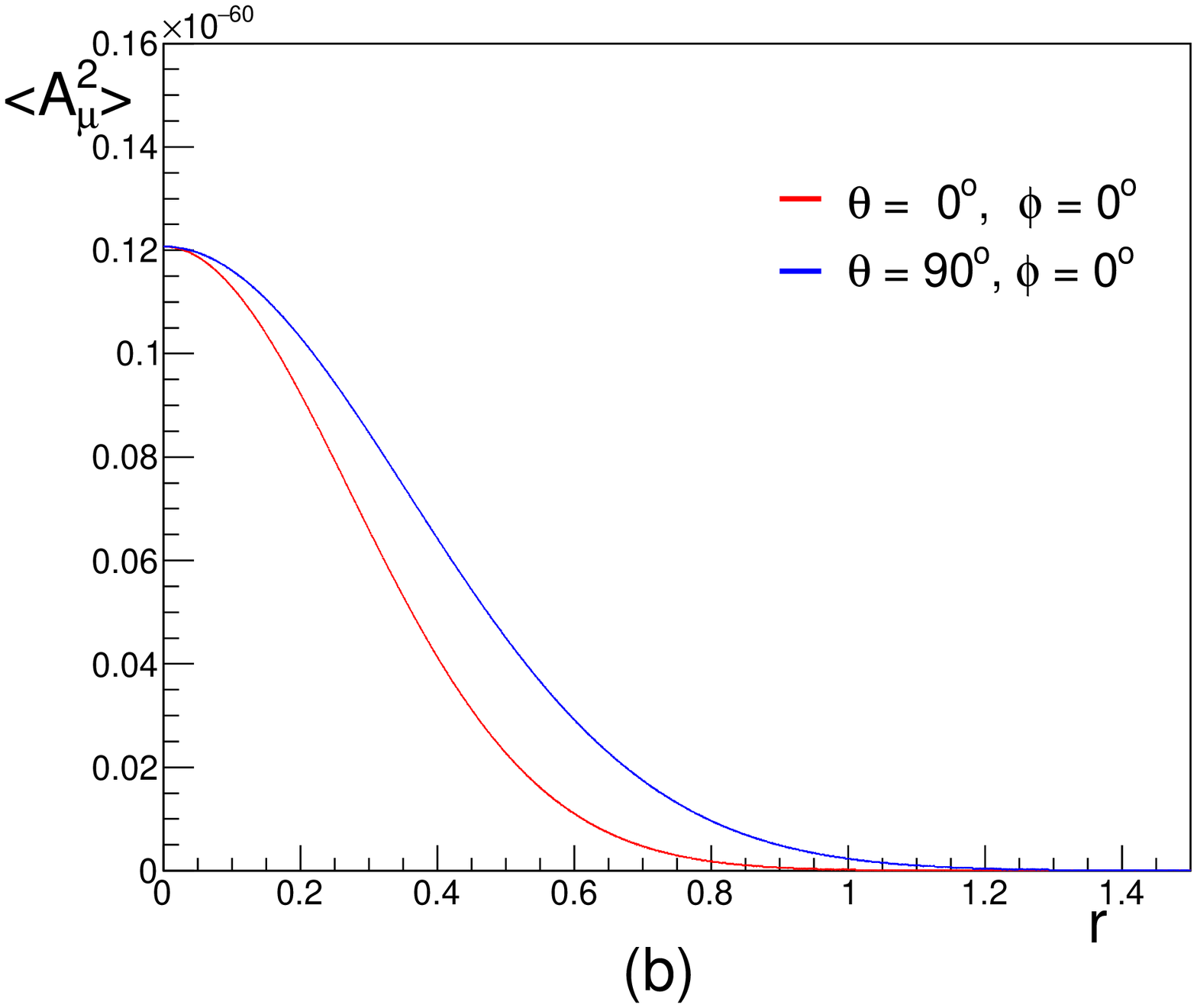}
 \caption{(Color online)
    (a) Equi-$\langle A_{\mu}^{2} \rangle$ surfaces for a nine quark configuration with quarks
      at $(0.0, 2.2, 0.0)$, $(-1.4, 1.7, 0.0)$, $(-2.2, 0.38, 0.0)$,
         $(-1.9, -1.1, 0.0)$, $(-0.75, -2.1, 0.0)$, $(0.75, -2.1, 0.0)$,
         $(1.9, -1.1, 0.0)$, $(2.2, 0.38, 0.0)$ and $(1.4, 1.7, 0.0)$.
    (b) $\langle A_{\mu}^{2} \rangle$ values along the two axes.
 }	
 \label{fig7}
 \end{figure}

\begin{figure}
 \includegraphics[width=0.76\linewidth]{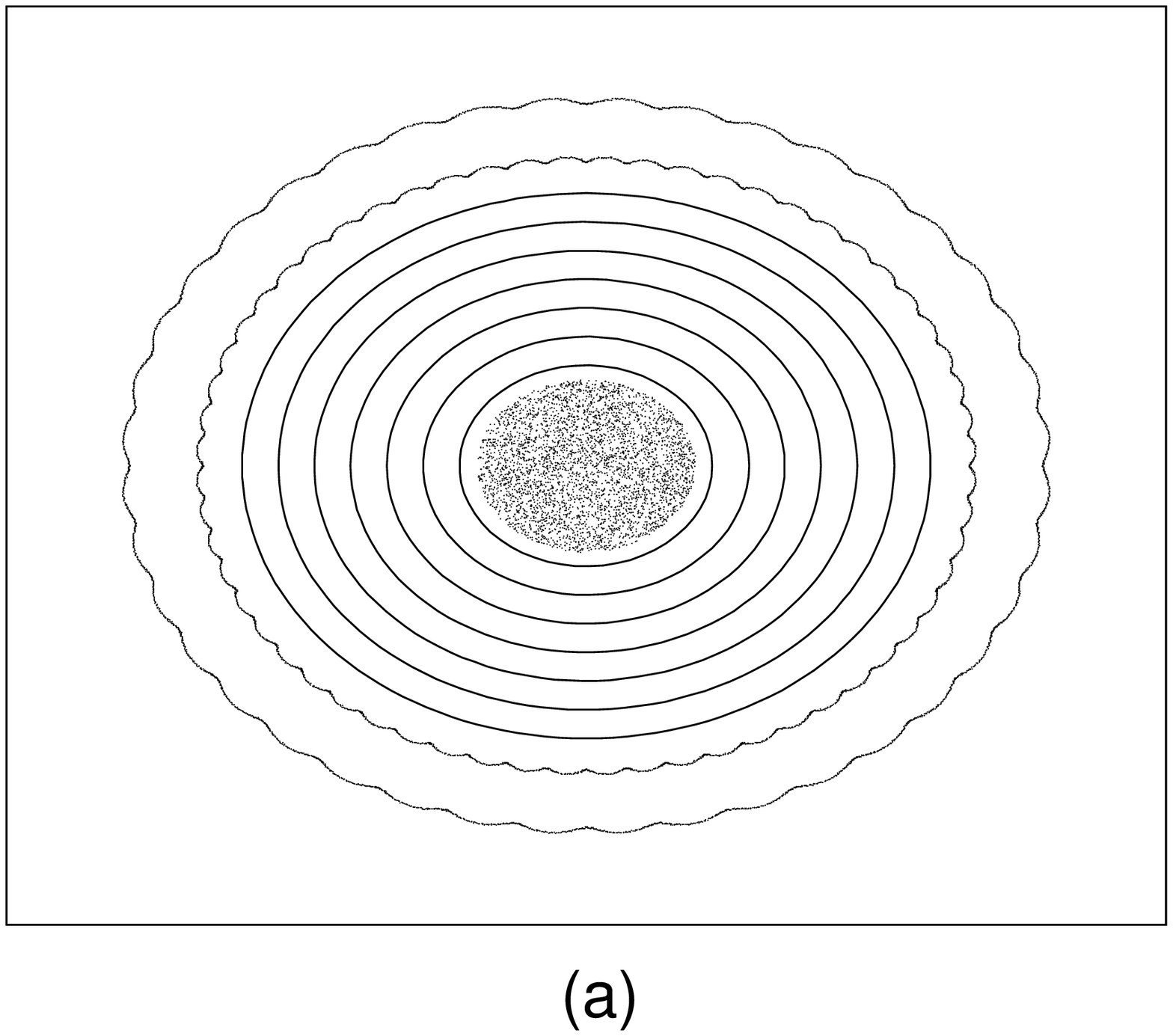}
 \includegraphics[width=0.76\linewidth]{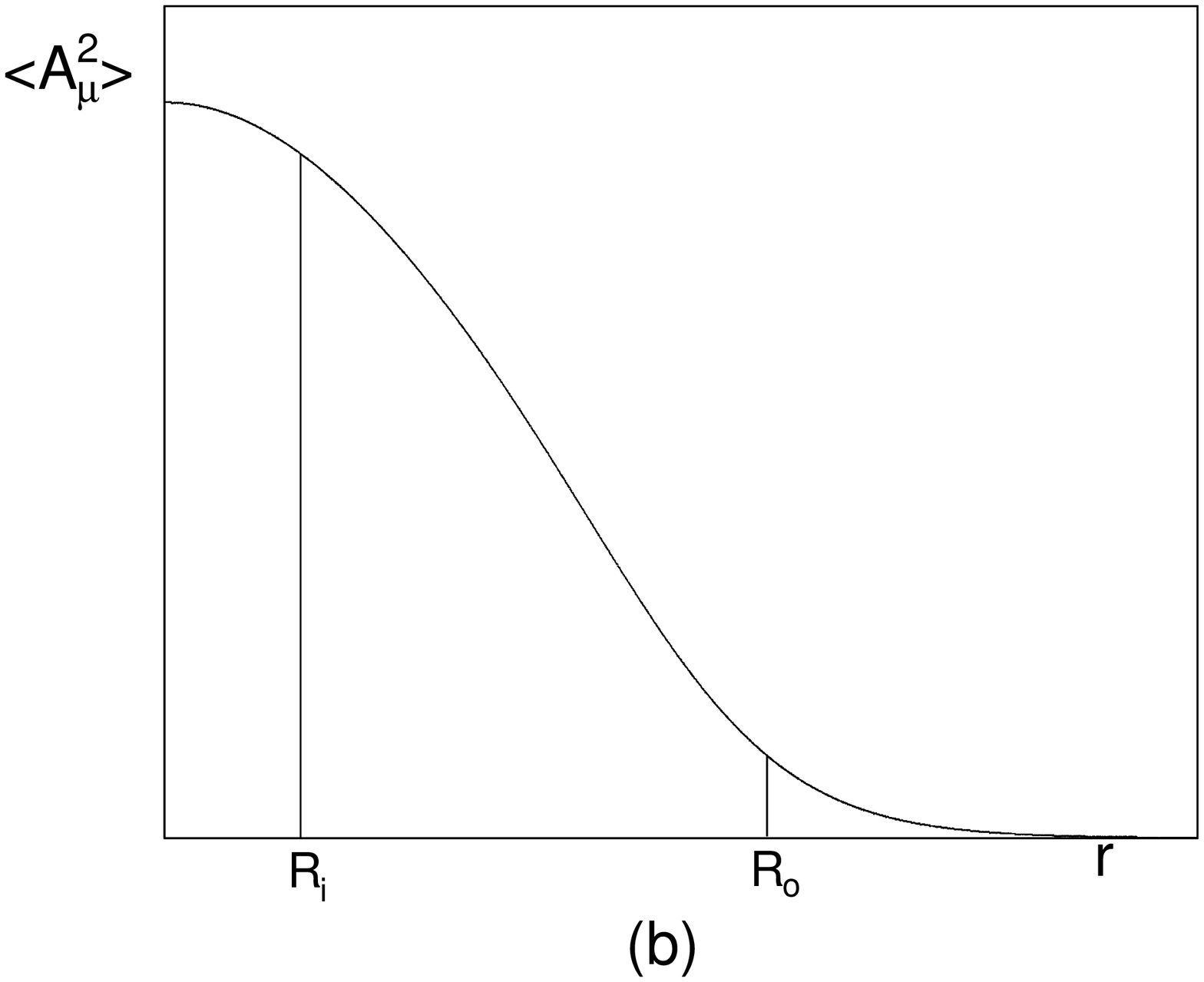}
 \caption{
   (a) Equi-$\langle A_{\mu}^{2} \rangle$ surfaces for a neutron star.
       The dotted region represents the quark-gluon plasma region
       and the wiggly lines represent the gauge slices outside of the neutron star.
   (b) Expected variation of $\langle A_{\mu}^{2} \rangle$ values from the center of
       the neutron star. $R_i$ is the radius for quark-gluon plasma region and
       $R_0$ is the radius of the neutron star.
  }	
 \label{fig8}
 \end{figure}

Similarly we can draw the configuration of a deuteron as in Fig.~\ref{fig5}(a)
and the condensate values are shown in Fig.~\ref{fig5}(b).
The six quarks are taken to be at the corners of a trigonal prism
and the condensate values are along the 3 lines.
%
%
%
%
The case of $^{3}\text{He}$ can be drawn as in Fig.~\ref{fig6}.
In Fig.~\ref{fig6}(a), the nine quarks are at the corners of three triangles,
and the condensate values are shown in Fig.~\ref{fig6}(b).
%
%
%
%
In Fig.~\ref{fig7}, another case with the nine quarks on a circle is shown.
We can find that the condensate values have maximum value at the center except for the positions
of quarks and decay exponentially just outside of the quark locations.
%
%
%
%
The same situation is expected to be valid for the case of quark locations
in the shape of spherical shell. By combining spherical shells, we can obtain
the case of sphere shape quark locations.
This case is approximately in the same situation as that of a neutron star,
but the neutron star contains the region of quark-gluon plasma.
For the quarks in the quark-gluon plasma region,
we cannot apply our nonperturbative formalism in a naive manner
but their contributions to long range confining region can be induced
from the above simple examples.
It is expected that in the quark-gluon plasma region the
$\langle A_{\mu}^{2} \rangle$ values become saturated due to the
perturbative interactions but the quark contributions to the
$\langle A_{\mu}^{2} \rangle$ values in the outer part of
the neutron star could be increased.
This situation is shown in Fig.~\ref{fig8}.
%
%
%
%
The radius for quark-gluon plasma region is represented by $R_i$ and
the radius of the whole neutron star is designated as $R_0$.
There exist no quarks outside of $R_0$ and because the
$\langle A_{\mu}^{2} \rangle$ value extends over the space
where no baryonic nor nuclear structures exist,
it is possible for low energy gluons to propagate along
the macroscopic spherical surfaces outside of neutron star.
These gluons are not observed by photons or by neutrinos
and therefore can be good candidates for dark matter.

The gluons propagating in free space may have originated from the Big Bang.
As the universe expands, the energies of gluons decrease below the value of
20~MeV and then they cannot be captured in baryons or nuclei.
However, they can interact with each other and form condensate structures
around neutron stars and even out to the scale of a galaxy.
If the sterile gluons left over after the big-bang nucleosynthesis existed
uniformly over the universe, they could exist in inter-galactic regions.
But somehow if they were gathering around galaxies, it is possible for them
to move from inter-galactic regions to intra-galactic regions.
The large scale condensate structures could be formed as nearly spherical
shapes centered at each centers of mass of galaxies.
Since they propagate along the spherical gauge slices, the mass effects
defined by Eq.(\ref{eq50}) could be related to the curvature scalar
and we get the radius dependences as
\begin{equation}
  m(r) \propto \frac{1}{r}.
\label{eq56}
\end{equation}
These dependences are consistent with the $\frac{1}{r^2}$ variation of the
dark matter density in a galaxy~\cite{R38}. For a crude estimation of dark matter density,
let's assume that the average energy of gluons propagating in intra-galactic region
is approximately 1~MeV fairly below the binding energy of deuteron 2.2~MeV.
If the average number of gluons per unit volume is assumed to be the same as
that of Big Bang photons 411~\cite{R39}, we have
\begin{equation}
  10^{-3} \times 411 \simeq 0.4\left(\text{GeV/cm$^3$}\right),
\label{eq57}
\end{equation}
which is just number obtained for the dark matter density in our galaxy.
%
%
%
%
\section{DISCUSSIONS}
\label{discussions}
%

In this paper, we reviewed the idea of generalized $\theta$ vacuum and
showed various structures of gauge fixed slices defined by the values
of dimension-2 condensates.
Because the gluons have to be quantized on these gauge slices,
it is essential to account for the curved structures of gauge slices.
The ordinary derivatives have to be replaced by covariant ones
and then from the second order form of the Lagrangian density
we get the curvature dependence of the gluon propagator.
The curvature dependences can be interpreted as mass effects
for the case of maximum symmetry and these mass effects might be
the newly found roles of gluons in hadrons and in any strongly
interacting systems as dark matter.

Gluonic contributions to hadronic masses are usually parametrized
as constituent quark masses which are very large compared with
the current quark masses determined by electromagnetically
or weakly interacting probes.
The differences can be taken as dark matter in the same sense of defining
the dark matter of a galaxy by the difference between measured
and observed masses of a galaxy. As for the large scale confinement
we can consider neutron stars as the centers for the large scale structures of gauge slices
and the structures can be extended even to the whole scale of a galaxy.
In general, at the center of a galaxy we expect the existence of supermassive black hole
and that black hole may serve as the center for the galaxial structure of gauge slices.
Since the curvature scalar decreases for the outer sphere from the center,
the mass density of dark matter can be explained systematically
from the large scale structure of gluon gauge slices.

Of course, there are many issues to be resolved with more detailed models.
First of all, the calculation of gluon propagator with more general form of
Ricci tensor is necessary.
This issue is closely related to the interpretation of curvature
in terms of the gluonic properties such as mass or other quantum numbers.
In this connection, the argument of gluon splitting~\cite{R40} could be exploited
because we need to separate parallel and transverse spin components to
follow up the changes along the curved gauge slices.
The determination of energy spectra of sterile gluons is another issue
and the coupling strength of very low energy gluons~\cite{R41} could affect
the final shape of cosmological dark matter distributions.

In hadronic scales, the calculations of hadronic masses with appropriate consideration
of gluonic effects are important issues. Traditionally an appropriate form of
confining potential is assumed between quarks and the spin-dependent
forces~\cite{R42} are derived by one gluon exchange diagrams in flat space.
There exist some divergence problems originating from masslessness
of the exchanged gluon and we need to resolve the problems with new propagator.
From the distribution of gauge slices it is also possible to deduce the form
of vacuum distribution functions which are used to define
nonperturbative quark propagator~\cite{R43}.
%
%
\section*{ACKNOWLEDGMENTS}
%
%
This work was supported by research funds of
Chonbuk National University in 2014.
%
%
%

%
%
\newpage
%
%
\newpage
\newpage
\newpage
\newpage
%
\newpage
%
\newpage
\newpage

\begin{references}
%

\bibitem{R1} A. A. Belavin {\it et al.}, Phys.~Lett. B~{\bf 59}, 85 (1975).
\bibitem{R2} G. 't Hooft, Phys.~Rev.~Lett.~{\bf 37}, 8 (1976);
             R. Jackiw and C. Rebbi, Phys.~Rev.~Lett.~{\bf 37}, 172 (1976).
\bibitem{R3} C. G. Callan, Jr., R. F. Dashen and D. J. Gross, Phys.~Lett. B~{\bf 63}, 334 (1976).
\bibitem{R4} S. J. Brodsky and R. Shrock,  Phys. Lett. B~{\bf 666}, 95 (2008).
\bibitem{R5} M. A. Shifman, A. I. Vainshtein and V. I. Zakharov, Nucl. Phys. B~{\bf 147}, 385 (1979);
             {\it ibid.} ~{\bf 147}, 448 (1979); {\it ibid.} ~{\bf 147}, 519 (1979).
\bibitem{R6} X. Li and C. M. Shakin, Phys. Rev. ~D~{\bf 70}, 114011 (2004);
             {\it ibid.} D~{\bf 71}, 074007 (2005).
\bibitem{R7}  E. J. Kim and J. B. Choi, J.~Korean~Phys.~Soc.~{\bf 61}, 1215 (2012).
\bibitem{R8}  D. E. Kharzeev, Nucl. Phys. A~{\bf 830}, 543c (2009).
\bibitem{R9}  F. V. Gubarev, L. Stodolsky and V. I. Zakharov, Phys.~Rev.~Lett.~{\bf 86}, 2220 (2001).
\bibitem{R10} D. Kharzeev, R. D. Pisarski, and M. H. G. Tytgat, Phys.~Rev.~Lett.~{\bf 81}, 512 (1998).
\bibitem{R11} B. I. Abelev {\it et al.} (STAR Collaboration), Phys.~Rev. C~{\bf 81}, 054908 (2010);
              B. Abelev {\it et al.} (ALICE Collaboration), Phys.~Rev.~Lett.~{\bf 110}, 012301 (2013).
\bibitem{R12} E. J. Kim, J. B. Choi and M. Q. Whang, J.~Korean~Phys.~Soc.~{\bf 56}, 1787 (2010).
\bibitem{R13} E. J. Kim {\it et al.}, J.~Korean~Phys.~Soc.~{\bf 58}, L1053 (2011).
\bibitem{R14} S. Weinberg, {\it Gravitation and Cosmology} (John Wiley \& Sons, New York, 1972), p103.
\bibitem{R15} Paticle Data Group, Chinese Phys. C~{\bf 38}, 090001 (2014), p353.
\bibitem{R16} Ph. Boucaud {\it et al.}, Phys.~Rev. D~{\bf 63}, 114003 (2001).
\bibitem{R17} Ph. Boucaud {\it et al.}, Phys.~Lett. B~{\bf 493}, 315 (2000).
\bibitem{R18} Ph. Boucaud {\it et al.}, Phys.~Rev. D~{\bf 66}, 034504 (2002).
\bibitem{R19} E. J. Kim and J. B. Choi, J.~Korean~Phys.~Soc.~{\bf 64}, L495 (2014).
\bibitem{R20} E. M. Kim {\it et al.}, J.~Korean~Phys.~Soc.~{\bf 64}, 1272 (2014).
\bibitem{R21} Ref.~\cite{R14}, p254.
\bibitem{R22} B. DeWitt, {\it 50 Years of Yang-Mills Theory} (World Scientific, edited by G. 't Hooft, 2005), p15.
\bibitem{R23} J. B. Choi and S. U. Park, J.~Korean~Phys.~Soc.~{\bf 24}, 263 (1991).
\bibitem{R24} S. V. Mikhailov and A. V. Radyushkin, Phys.~Rev. D~{\bf 45}, 1754 (1992).
\bibitem{R25} X. Li and C. M. Shakin, Phys. Rev. ~D~{\bf 71}, 074007 (2005).
\bibitem{R26} See the parameters in Ref.~\cite{R25}.
\bibitem{R27} Ref.~\cite{R21}, p140.
\bibitem{R28} Ref.~\cite{R21}, p383.
\bibitem{R29} K. I. Kondo and T. Shinohara, Phys. Lett. B~{\bf 491}, 263 (2000).
\bibitem{R30} J. J. J. Kokkedee, {\it The Quark Model} (W. A. Benjamin, New York, 1969), p15.
\bibitem{R31} M. Takizawa, K. Kubodera and F. Myhrer, Phys.~Lett. B~{\bf 261}, 221 (1991).
\bibitem{R32} J. Kuti and V. F. Weisskopf, Phys. Rev. ~D~{\bf 4}, 3418 (1971).
\bibitem{R33} S. F. Tuan {\it et al.}, Phys.~Rev. D~{\bf 10}, 2124 (1974).
\bibitem{R34} Ref.~\cite{R15}, p732.
\bibitem{R35} S. J. Brodsky {\it et al.}, arXiv:1005.4610 [nucl-th].
\bibitem{R36} F. Weber {\it et al.}, {\it Neutron Stars and Pulsars} (Cambridge Univ. Press, edited by J. van Leeuwen, 2013), p61.
\bibitem{R37} Ref.~\cite{R15}, p78.
\bibitem{R38} See Ref.~\cite{R15}.
\bibitem{R39} Ref.~\cite{R15}, p369.
\bibitem{R40} A. J. Niemi and N. R. Walet, Phys.~Rev. D~{\bf 72}, 054007 (2005).
\bibitem{R41} C. S. Fischer, A. Maas and J. M. Pawlowski, arXiv:0810.1987 [hep-ph].
\bibitem{R42} E. Eichten and F. Feinberg, Phys.~Rev. D~{\bf 23}, 2724 (1981);
              D. Gromes, Phys. Rep.~{\bf 200}, 186 (1991).
\bibitem{R43} See, for example, O. Andreev, Phys.~Rev. D~{\bf 82}, 086012 (2010).
%
%
\end{references}
\end{document}